\documentclass[a4paper]{spie}  

\usepackage{amsmath,amsfonts,amssymb}
\usepackage{graphicx}
\usepackage[colorlinks=true, allcolors=blue]{hyperref}

\newcommand{\fesc}{$f_{esc}$\ }

\title{CUBES, the Cassegrain U-Band Efficient Spectrograph.}

\author[a]{S. Cristiani}
\author[m]{J. M. Alcalá}
\author[r]{S. H. P. Alencar}
\author[s]{S. A. Balashev}
\author[t,u]{N. Bastian}
\author[b]{B. Barbuy}
\author[v]{U. Battino}
\author[o]{A. Calcines}
\author[a]{G. Calderone}
\author[e]{P. Cambianica}
\author[n]{R. Carini}
\author[ar]{B. Carter}
\author[ab,ac]{S. Cassisi}
\author[l]{B. V. Castilho}
\author[k]{G. Cescutti}
\author[j]{N. Christlieb}
\author[a]{R. Cirami}
\author[a]{I. Coretti}
\author[d]{R. Cooke}
\author[c]{S. Covino}
\author[e]{G. Cremonese}
\author[ad,ae]{K. Cunha}
\author[a]{G. Cupani}
\author[g]{A.~R.~da Silva}
\author[m]{V. De Caprio}
\author[ag]{A. De Cia}
\author[x]{H. Dekker}
\author[af]{V. D’Elia}
\author[at]{G. De Silva}
\author[b]{M.Diaz}
\author[a]{P. Di Marcantonio}
\author[m]{D. D'Auria} 
\author[a]{V. D'Odorico}
\author[ai]{A. Fitzsimmons }
\author[b]{H. Ernandes}
\author[p]{C. Evans}
\author[a]{M. Franchini}
\author[c]{M. Genoni}
\author[q]{B. G\"ansicke}
\author[g]{R.~E.~Giribaldi}
\author[l]{C. Gneiding}
\author[e]{A. Grazian}
\author[ah]{C. J. Hansen}
\author[ak]{F. La Forgia}
\author[c]{M. Landoni}
\author[ak]{M. Lazzarin}
\author[i]{D. Lunney}
\author[b]{W. Maciel}
\author[al]{W. Marcolino}
\author[m]{M.Marconi}
\author[am]{A. Migliorini}
\author[i]{C. Miller}
\author[an,ao]{P. Noterdaeme}
\author[h]{C. Opitom}
\author[c]{G. Pariani}
\author[g]{B.~Pilecki}
\author[n]{S. Piranomonte}
\author[j]{A. Quirrenbach}
\author[c]{E.M.A. Redaelli}
\author[ad]{C. B. Pereira}
\author[aq]{S. Randich}
\author[b]{S. Rossi}
\author[i]{R. Sanchez-Janssen}
\author[j]{W. Seifert}
\author[g]{R. Smiljanic}
\author[h]{C. Snodgrass}
\author[j]{I. Stilz}
\author[j]{J. St\"urmer}
\author[f]{E. Vanzella}
\author[n]{P. Ventura}
\author[l]{O. Verducci}
\author[i]{C. Waring}
\author[i]{S. Watson}
\author[i]{M. Wells}
\author[ar]{D. Wright}
\author[at]{T. Zafar}
\author[c]{A. Zanutta}
\affil[a]{INAF - Osservatorio Astronomico di Trieste, via G. B. Tiepolo 11, 34131 Trieste, Italy}
\affil[b]{Universidade de São Paulo, IAG, Rua do Matão 1226, São Paulo, Brazil}
\affil[c]{INAF - Osservatorio Astronomico di Brera, via E. Bianchi 46, 23807 Merate, Italy}
\affil[d]{Centre for Extragalactic Astronomy, Durham University, Durham DH1 3LE, UK}
\affil[e]{INAF - Osservatorio Astronomico di Padova, Vicolo dell'Osservatorio 3, Padova, Italy}
\affil[f]{INAF - Osservatorio di Astrofisica e Scienza dello Spazio, Bologna, Italy}
\affil[g]{Nicolaus Copernicus Astronomical Center, Polish Academy of Sciences, Warsaw, Poland}
\affil[h]{Institute for Astronomy, University of Edinburgh, Royal Observatory, Edinburgh, UK}
\affil[i]{STFC - United Kingdom Astronomy Technology Centre (UK ATC), Edinburgh, UK}
\affil[j]{Landessternwarte, Zentrum für Astronomie der Universität Heidelberg, Heidelberg, Germany}
\affil[k]{Dipartimento di Fisica, Sezione di Astronomia, Università di Trieste, Italy}
\affil[l]{LNA/MCTI, Rua Estados Unidos, 154 - 37504-364, Itajubá, Brazil}
\affil[m]{INAF-Osservatorio Astronomico di Capodimonte, via Moiariello 16, 80131 Napoli, Italy}
\affil[n]{INAF - Osservatorio Astronomico di Roma, Via Frascati 33,
Monte Porzio Catone, Italy}
\affil[o]{Durham University,  Centre for Advanced Instrumentation, Department of Physics, UK}
\affil[p]{European Space Agency (ESA), ESA Office, Space Telescope Science Institute, 3700 San Martin Drive, Baltimore, MD 21218, USA}
\affil[q]{University of Warwick, Department of Physics, Gibbet Hill Road, Coventry, CV7 4AL, UK}
\affil[r]{Departamento de Fisica - ICEx - UFMG, Belo Horizonte, MG, Brazil}
\affil[s]{Ioffe Institute, Polyteknicheskaya 26,  HSE University, 194021 Saint-Petersburg, Russia}
\affil[t]{Donostia International Physics Center (DIPC), Donostia-San Sebastián, Guipuzkoa, Spain}
\affil[u]{IKERBASQUE Basque Foundation for Science, E-48013 Bilbao, Spain }
\affil[v]{University of Hull, E.A. Milne Centre for Astrophysics, Hull, UK}
\affil[x]{Consultant Astronomical Instrumentation, Alpenrosenstr.15, 85521 Ottobrunn, Germany}
\affil[ab]{INAF - Osservatorio Astronomico di Abruzzo, Via M. Maggini, I-64100 Teramo, Italy}
\affil[ac]{INFN - Sezione di Pisa, Largo Pontecorvo 3, I-56127 Pisa, Italy}
\affil[ad]{Observat\'orio Nacional, Rua Gen. Jos\'e Cristino 77, S\~ao Crist\'ov\~ao, Rio de Janeiro, Brazil }
\affil[ae]{Steward Observatory, University of Arizona, 950 N. Cherry Ave., Tucson, AZ, 85719}
\affil[af]{Italian Space Agency - Space Science Data Centre, via del Politecnico snc, Rome, Italy}
\affil[ag]{Department of Astronomy, University of Geneva, Chemin Pegasi 51, Versoix, Switzerland}
\affil[ah]{Goethe University Frankfurt, Institute for Applied Physics, Frankfurt am Main, Germany}
\affil[ai]{Astrophysics Research Centre, Queen's University, Belfast, UK}
\affil[ak]{Dipartimento di Fisica e Astronomia dell'Università, Vicolo dell'Osservatorio 3, Padova, Italy}
\affil[al]{Observat\'orio do Valongo, Universidade Federal do Rio de Janeiro, Brazil}
\affil[am]{INAF - Institute of Space Astrophysics and Planetology, Roma, Italy}
\affil[an]{Franco-Chilean Laboratory for Astronomy, Las Condes, Santiago, Chile}
\affil[ao]{Institut d’Astrophysique de Paris, CNRS-SU, UMR 7095, 98bis bd Arago, Paris, France}
\affil[aq]{INAF - Osservatorio Astrofisico di Arcetri, Largo E. Fermi 5, 50125, Firenze, Italy}
\affil[ar]{Centre for Astrophysics, University of Southern Queensland, Toowoomba 4350, Australia}
\affil[at]{Australian Astronomical Optics, Macquarie University, North Ryde, NSW 2113, Australia}

\authorinfo{Further author information: (Send correspondence to Stefano Cristiani)\\Stefano Cristiani: E-mail: stefano.cristiani@inaf.it, 
\url{https://wwwuser.oats.inaf.it/cristiani/} \\ }

\begin{document} 
\maketitle

\begin{abstract}

In the era of Extremely Large Telescopes, the current generation of 8-10m facilities are likely to remain competitive at ground-UV wavelengths for the foreseeable future. 
The Cassegrain U-Band Efficient Spectrograph (CUBES) has been designed to provide high-efficiency ($>40$\%)  observations in the near UV (305-400 nm requirement, 300-420 nm goal) at a spectral resolving power of $R>20,000$ 
(with a lower-resolution, sky-limited mode of $R \sim 7,000$).
With the design focusing on maximizing the instrument throughput  (ensuring a Signal to Noise Ratio (SNR)$\sim 20$ per high-resolution element at 313 nm for $U \sim 18.5$\,mag objects in 1h of observations), it will offer new possibilities in many fields of astrophysics, providing access to key lines of stellar spectra: a tremendous diversity of iron-peak and heavy elements, lighter elements (in particular Beryllium) and light-element molecules (CO, CN, OH), as well as Balmer lines and the Balmer jump (particularly important for young stellar objects). The UV range is also critical in extragalactic studies: the circumgalactic medium of distant galaxies, the contribution of different types of sources to the cosmic UV background, the measurement of H$_2$ and primordial Deuterium in a regime of relatively transparent intergalactic medium, and follow-up of explosive transients.
The CUBES project completed a Phase A conceptual design in June 2021 and has now entered the detailed design and construction phase. First science operations are planned for 2028.

\end{abstract}


\section{INTRODUCTION}
\label{sec:intro}  
The four 8.2m telescopes of the Very Large Telescope (VLT) at the European Southern Observatory (ESO)
are the world’s most scientifically productive groundbased observatory in the visible and infrared. Looking to the future of the VLT there is a long-standing aspiration for an optimised ultraviolet (UV) spectrograph\cite{Barbuy2014}.

The European Extremely Large Telescope (ELT), under construction in northern Chile by ESO, with a primary aperture of 39m will be unprecedented in its light-gathering power, coupled with exquisite angular resolution via correction for atmospheric turbulence by adaptive optics (AO). At variance to current large telescopes such as VLT, AO is an integral part of the ELT, which has a novel five-mirror design including a large adaptive mirror (M4) and a fast tip-tilt mirror (M5). The choice of protected silver (Ag+Al) for the ELT mirror coatings (excl. M4) ensures a durable, proven surface with excellent performance across a wide wavelength range, but the performance drops significantly in the blue-UV part of the spectrum compared to bare aluminium. ESO is actively researching alternative coatings, but in the short-medium term we can assume that the performance of the ELT in the blue-UV will be limited in this regard. Indeed, during the Phase A study of the MOSAIC multi-object spectrograph\cite{Evans2016} it was concluded that a blue-optimised instrument on the VLT could potentially be competitive with the ELT at wavelengths shorter than 400 nm.
Motivated by this, in 2018 we revisited \cite{Evans2018} the Phase A study undertaken in 2012 of the Cassegrain U-band Brazilian-ESO Spectrograph. The past study investigated a $R\sim 20$k spectrograph operating at ‘ground UV’ wavelengths (spanning 300-400 nm) to open-up exciting new scientific opportunities compared to the (then) planned instrumentation suite for Paranal Observatory\cite{Barbuy2014,Bristow2014}. 

In January 2020 ESO issued a Call for Proposal for a Phase A study of a UV Spectrograph to be installed at a Cassegrain focus of the VLT, with the goal of high-efficiency ($>40$\%) and intermediate resolving power ($\sim$ 20K) in the ground-UV domain (305-400 nm requirement, 300-420 nm goal). 
In May 2020 the CUBES (Cassegrain U-Band Efficient Spectrograph) Consortium, led by INAF, was selected to carry out the study. 
The CUBES project completed a Phase A conceptual design in June 2021.
After the endorsement by the ESO Council at the end of 2021,
Phase B started in February 2022 with the signature of the Construction Agreement between ESO and the leading institute of the CUBES Consortium, opening the detailed design and construction phase.
Here we report the present status of the project, which will provide a world-leading UV capability for ESO from 2028 well into the ELT era.  
\section{SCIENCE WITH CUBES}
\label{sec:science}  
The CUBES science case spans a broad range of contemporary astrophysics. Here we highlight key cases across Solar System, Galactic, and extragalactic science that are driving the design.

\subsection{Searching for water in the asteroid belt\cite{Opitom2022}}
The search for water in our solar system is far from complete and is tremendously difficult with ground-based facilities given the water content of Earth's atmosphere.  The most powerful ground-based probe of water outgassing from small bodies of the solar system is OH emission at 308 nm. Observation of the OH line is only possible with current facilities for a few active comets while they are near the Sun and Earth. This severely limits studies of water production around their orbits and we miss seasonal effects that the Rosetta mission revealed to be important. It also prevents study of most comets as they are simply too faint. Even more tantalising are studies of main-belt comets – bodies in asteroidal orbits that undergo activity (usually detected by a dust tail or trail) that is thought to arise from sublimation (e.g. recurrent activity near perihelion). Constraining the OH emission of such objects is well beyond our current capabilities. Main-belt comets have typical sizes that are very common in the asteroid belt, so detection of outgassing water with CUBES would point to a potentially large population of icy bodies, hence a large reservoir of water, of considerable interest in models of the formation and evolution of the inner solar system.
\begin{figure}
    \centering
    \includegraphics[width=0.585\textwidth]{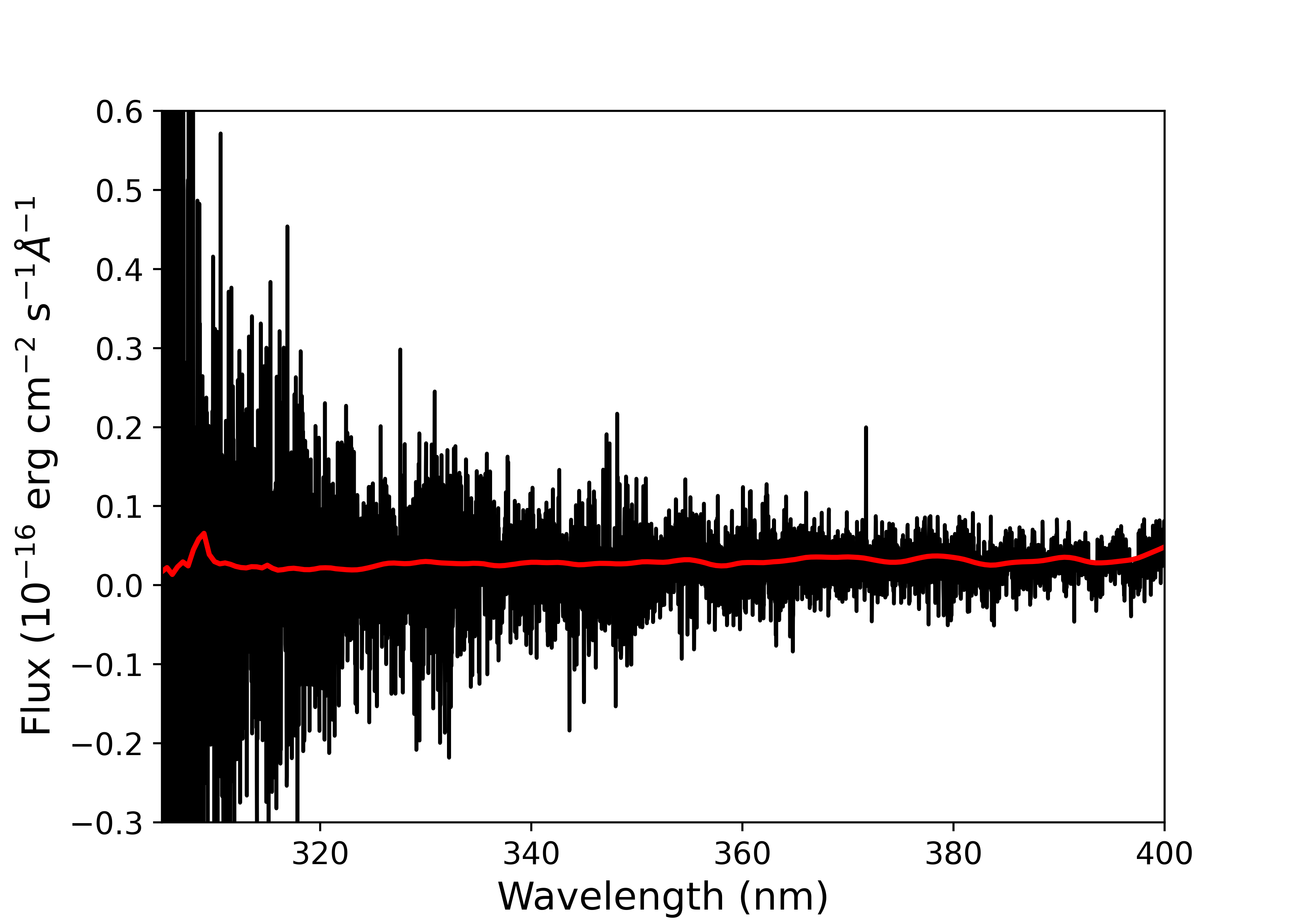}
    \includegraphics[width=0.405\textwidth]{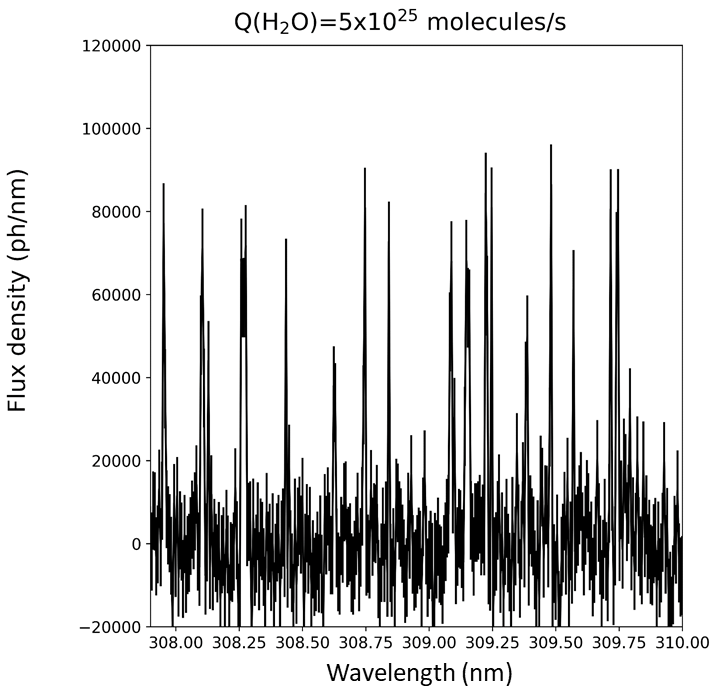}
    \caption{{\it Left}: X-Shooter spectrum of main belt comet P/2012
T1 (black) from \cite{Snodgrass2017} . The red curve is an artificial spectrum
of P/2012 T1 produced with the Planetary Spectrum Generator
\cite{Villanueva2018} for a water production rate of $5 \times 10^{25}$ molecules/s. Significantly
better signal-to-noise is needed in the near UV to search for very weak outgassing of OH
by main belt comets. {\it Right}: Simulated extracted spectrum of main belt comet P/2012 T1 with CUBES around the peak of the
OH band for water production rates of $5 \times 10^{25}$
molecules/s. The simulated comet spectrum
was produced using the Planetary Spectrum Generator \cite{Villanueva2018} combined with the CUBES end-to-end simulator \cite{Genoni2022}.}
    \label{fig:Comets}
\end{figure}
\subsection{Accretion, winds \& outflows in YSOs}
Studying the evolution of circumstellar disks, mass accretion around young stars, as well as outflow and winds is essential to fully understand the formation of protoplanets. This requires multi-wavelengths studies of stars during the first 10\,Myr of their evolution and in particular Classical T Tauri stars (CTTS). They are young, low- to solar-mass stars that are actively accreting mass from planet-forming disks. Spectroscopic surveys of CTTS in nearby star-forming regions have been used to study the mutual relationships between accretion, jets and disk structure. CUBES, with its increased UV sensitivity and coverage of a critical wavelength range will enable more detailed studies of the accretion processes/wind-outflows than currently possible as well as studies of CCTS in star-forming region at larger distances, for example. 

\subsection{Bulk composition of exo-planets} 
In the past 25 years we have gone from not knowing if the Solar System is unique to understanding that it is normal for stars to have planets. Characterising exo-planets and understanding their formation and evolution are major research fields, but the fundamental question of what these other worlds are made of cannot be answered from studies of planets orbiting main-sequence stars. 
The only approach available at present is estimating the bulk composition of exo-planet systems from spectroscopic analysis of evolved white dwarf (WD) stars accreting debris from tidally-disrupted planetesimals. WDs are hot so most of their abundance diagnostics are in the near-UV (e.g. Sc, Ti, V, Cr, Mn, Fe, Ni). However, WDs are intrinsically faint, and only about twenty systems currently have precise abundances so far. CUBES will transform this exciting area of exo-planet research by increasing the sample of known exo-planetesimal compositions to a level comparable to the number of meteorite samples with abundances, thus providing precise constraints on the next generation of planet-formation models.

\subsection{Stellar nucleosynthesis}
The origin of the elements in the periodic table has a prominent role in both astronomy and nuclear physics. Indeed, production of the chemical elements that we are made of and use every day in our lives is one of the most profound questions we can ask. Each chemical isotope has a complex production channel, which includes numerous nuclear reactions. Remarkably, the spectral features of more than a quarter of the chemical elements are only observable in the near UV, but the low efficiency of instruments in this domain severely restricts the scope of current studies. To calibrate and test theoretical predictions of stellar evolution and nucleosynthesis we require high-resolution, near-UV spectroscopy of a much larger number and wider diversity of stars. Three showcases motivate new observations in this very broad and active field.
\subsubsection{Metal-poor stars and light elements}
 A key case is to probe the early chemical evolution of the Galaxy, via chemical abundance patterns in the oldest, low-mass stars that formed from material enriched by the first supernovae. The so-called Carbon-enhanced metal-poor (CEMP) stars are the perfect probes to investigate nucleosynthesis by the first stars. Ongoing imaging surveys are finding more examples, but we lack the ground-UV sensitivity to obtain spectra for all but the very brightest. CUBES will enable quantitative spectroscopy for large samples of metal-poor stars, providing direct estimates for a broad range of heavy elements, as well as valuable constraints on CNO elements. The abundances of these light elements in metal-poor stars bring a wealth of information on stellar evolution and the chemical evolution of the Galaxy. The near-UV contains strong bands of CN,  OH molecular bands at $< 320$ nm and the unique NH band at  336 nm which is the only feature allowing to derive nitrogen abundances directly.

\subsubsection{Heavy-element nucleosynthesis}
Stellar abundances from CUBES will provide critical tests of the various production channels of heavy elements for both r- and s-process elements. Determining the abundances of neutron-capture elements in metal-poor stars is fundamental to understand the physics of these processes and the chemical evolution of the Galaxy as well as the origin of the Galactic halo. Abundances of at least one neutron-capture element (typically Ba) have been measured in a significant number of stars in the Galactic halo and more will be measured by upcoming instruments. However, a single element is not sufficient to gain a clear understanding of the different processes. All of the key elements have to be probed. Since many of these are in the UV domain (e.g. Hf, Sn, Ag, Bi) and have only been measured for a very restricted number of stars, CUBES will play a critical role to fill this gap.
\begin{figure}[hb]
    \centering
    \includegraphics[width=0.45\textwidth]{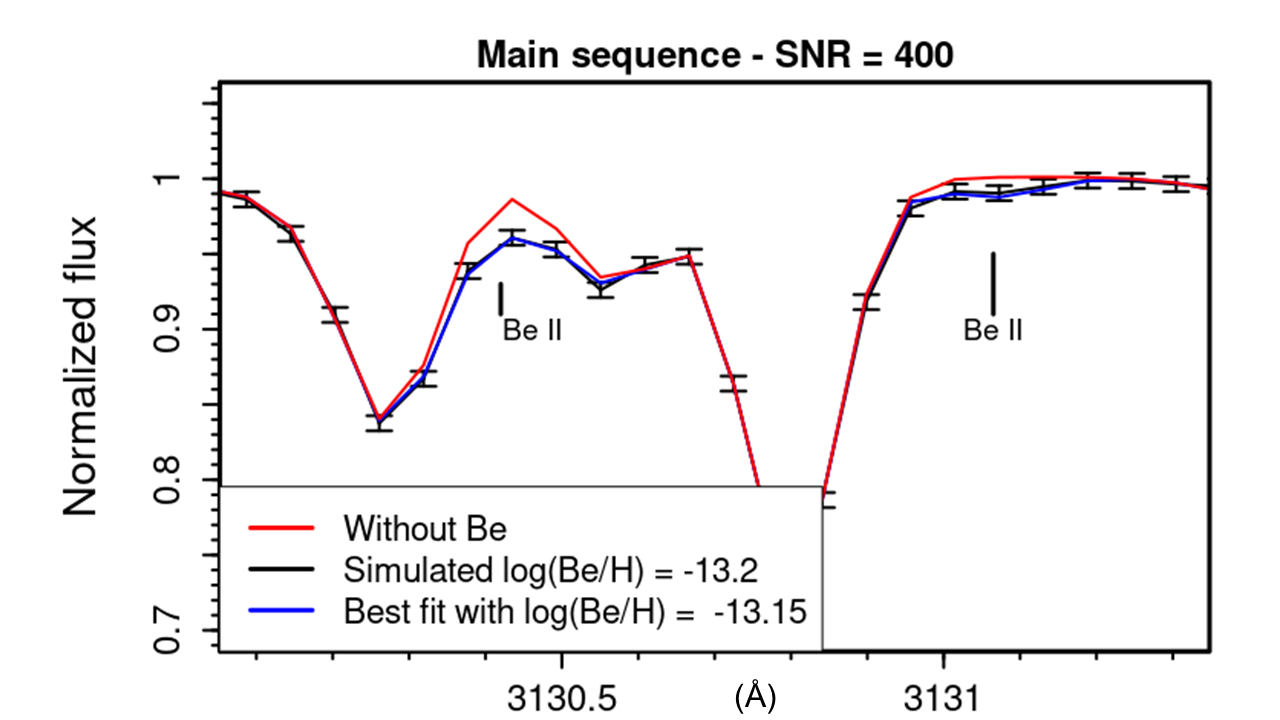}
    \includegraphics[width=0.265\textwidth]{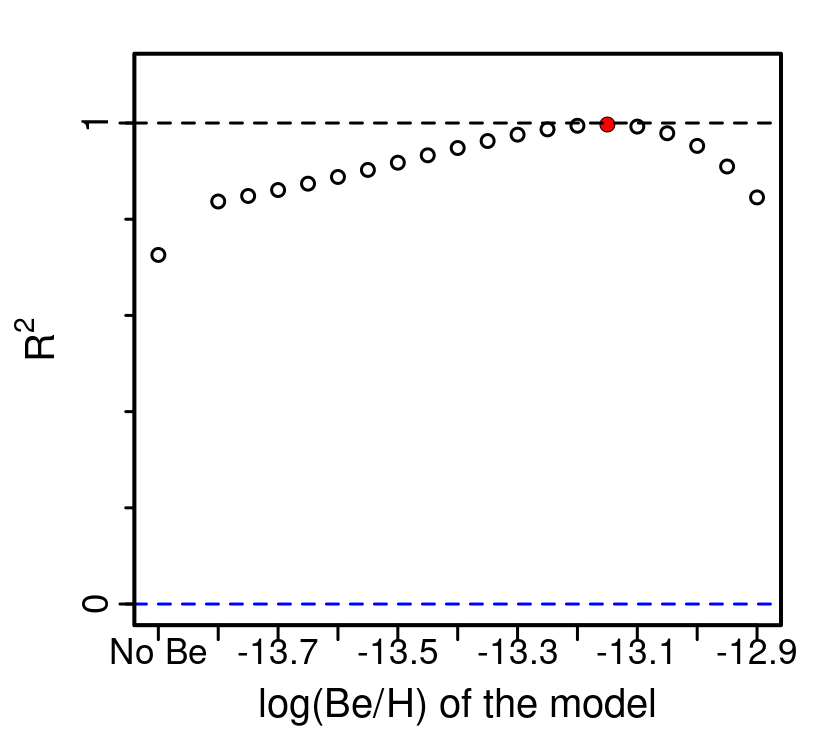}
    \includegraphics[width=0.265\textwidth]{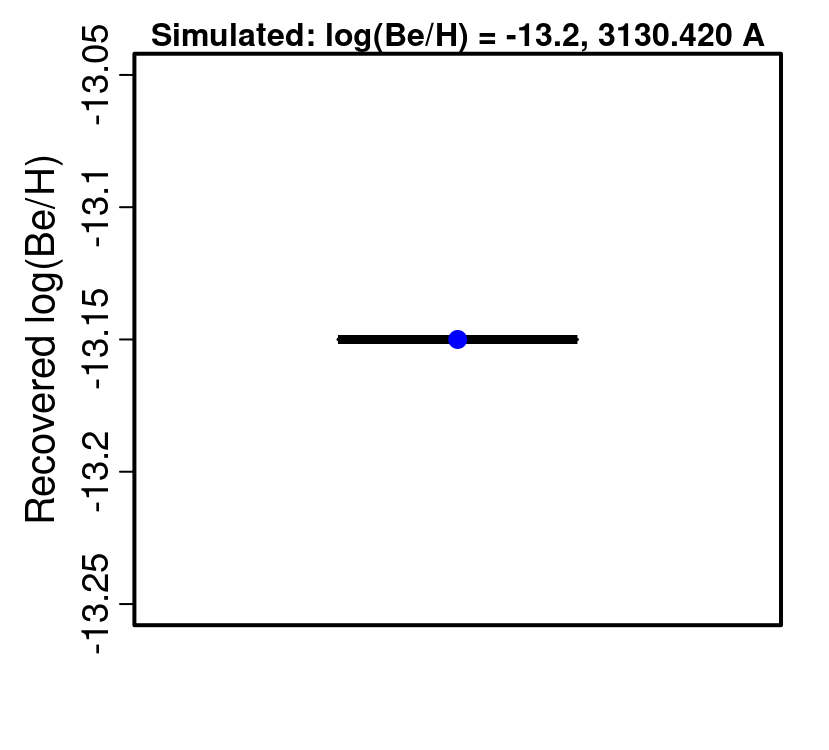}
    \caption{Results of fitting five pixels around the 3130 \AA\ line in the case of a bright main-sequence star with [Fe/H] = $-$3.0 and input abundance of $\log(\mathrm{Be/H})$ = $-$13.0, with SNR = 400. {\it Left:} One selected best fit with $\log$(Be/H) = $-$13.15 for the observation that was simulated with input abundance of $\log$(Be/H) = $-$13.20. {\it Center:} The coefficient of determination, $R^2$, values obtained when fitting the Be line with spectra with a range of abundances. {\it Right:} Boxplot of the best fitting abundances \cite{Smiljanic2022}.}
    \label{fig:bright.ms.3130.m3.be13p2}
\end{figure}
\subsubsection{Beryllium}
Although one of the lightest, simplest elements, questions remain on the production of Be in the early Universe. Recent results are consistent with no primordial production, but larger samples are required to investigate this further. Moreover, $^9$Be is never produced by stellar evolution (only destroyed in stellar interiors), so it has been suggested that it might be used as a cosmic clock to trace star-formation histories (e.g. and in this it would nicely complemente results from Gaia). Only ~200 stars have Be abundances so far (limited to $V$ $\sim$ 12 mag in a few hrs with UVES\cite{UVES2000}). CUBES will provide large homogeneous samples of Be abundances in stars belonging to different populations up to three magnitudes deeper (e.g.~follow-up of metal-poor stars from the 4MOST survey), providing new insights into its production and tracing the star-formation history of the Galaxy. An example of the expected instrument performance in the context of the measurement of Be abundances is shown in Fig.~\ref{fig:bright.ms.3130.m3.be13p2}.
\subsection{Primordial deuterium abundance}
Within the Standard Model of particle physics and cosmology we ‘know’ the content of the Universe 
to within a few percent. However, there is still no accepted model for dark energy and dark matter, or why the Universe contains baryons instead of antibaryons, or even why the Universe contains baryons at all. We are also missing crucial properties of neutrinos (e.g. their hierarchy, why they change flavour, and the number that existed during the earliest phases of the Universe). These limitations indicate there is missing physics in our understanding. Thus, new cosmological observations may allow us to discover something fundamental about the early properties of the Universe.
Some of these questions can be investigated by measuring the nuclides produced a few minutes after the Big Bang. The primordial deuterium (D/H) abundance is currently our most reliable probe of Big Bang Nucleosynthesis as it can be deduced to a precision of $\sim 1$\% using near-pristine gas clouds that imprint DI and HI absorption lines on spectra of background quasars. This is painstaking observational work, with only seven systems with precise D/H estimates so far, most of which are at redshifts of 2.8 to 3.1. CUBES will provide a large, reliable sample of D/H estimates. Its significant gain at the shortest wavelengths compared to existing facilities will enable observations at lower redshifts (less contamination by the Lyman-alpha forest) giving more absorption-line systems from which to estimate D/H and smaller uncertainties. Moreover, we already know suitable quasar sightlines that intersect near-pristine gas clouds at redshifts $2.3 < z < 2.8$ with the desirable properties (e.g. quiescent, near-pristine, mostly neutral) to accurately infer the primordial D/H ratio.
\begin{figure}[hb]
    \centering
    \includegraphics[width=0.85\textwidth]{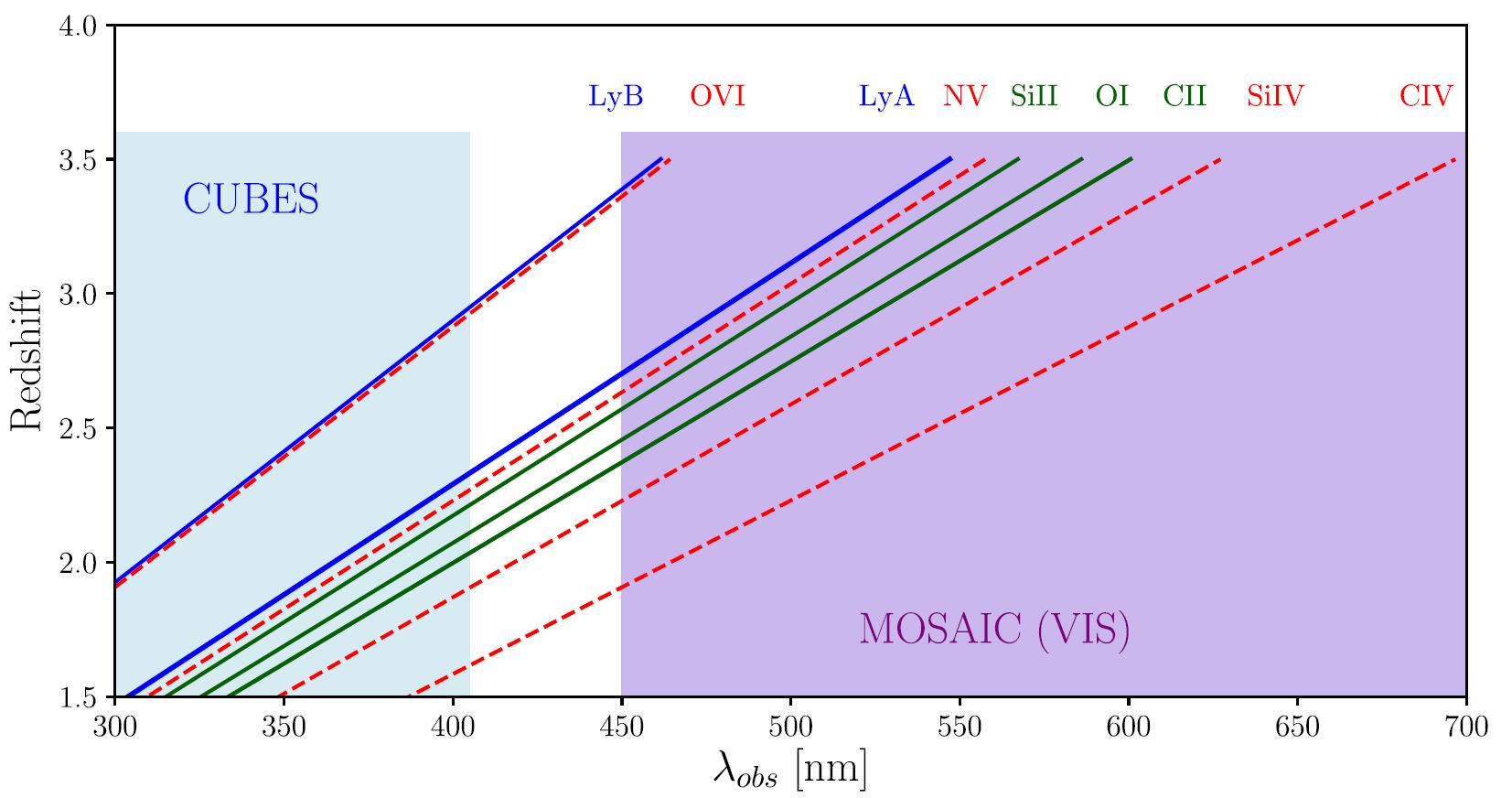}
     \caption{Lines available in the ground-UV for studies of the CGM of galaxies at $1.5 < z < 2.3$ (or $1.9 < z < 2.9$ for O VI absorbers). They span a range of ionisation states (green = low, red = high) to study the relative fractions of cold vs. warm vs. hot gas, at redshifts where the contamination of key ions such as O VI by the Lyman forest is less severe. CUBES observations will neatly complement future observations in the blue-visible with the ELT (Phase A design of) MOSAIC\cite{MOSAIC_2020}, which is targeting the CGM in galaxy halos at $z \sim 3$}
    \label{fig:lines_z_range}
\end{figure}
\subsection{The missing baryonic mass at the cosmic noon}
For almost thirty years scientists have been looking for a definitive answer to the question “Where are the baryons?” (e.g. \cite{Fukugita1998}). Progress has been made possible 
at low redshifts by the dispersion measure in FRBs \cite{Macquart2020} and at $z > 1.5$ by observations and simulations of the Lyman forest (e.g. \cite{Weinberg1997}). However it is unclear how baryonic matter is distributed among the different gaseous components and we have still to understand and constrain the mechanisms (stellar and AGN feedback, accretion, etc.) that determined the observed distribution. A UV efficient spectrograph with relatively high resolution offers the possibility to dig into the complex nature of the inter- and circum-galactic gas at $z \sim$ 1.5 to 3, via two experiments with quasar absorption lines: 
\begin{enumerate}
    \item The baryons in the diffuse IGM are chased through the detection and analysis of Lyman-$\alpha$ lines at $z \simeq 1.5$ to 2.3. This redshift range, immediately after the era of peak star-formation in the Universe, is poorly investigated due to observational difficulties (namely the low efficiency of ground-based spectrographs in the UV) but is critical to connect the low- and high-redshift results. The relatively low number density of lines in the Lyman forest allows us to minimise, even with observations at $R \sim 20K$, the impact of blending with metal lines (which is not the case at larger redshifts).
\item Observing the O VI absorption lines at $1.9 < z < 2.9$ to trace the warm-hot gas at $T > 10^5$ K, associated with the IGM or with the CGM\cite{Lehner2014}.
\end{enumerate}
In both cases spectroscopy in the wavelength range at $\lambda > 400$ nm is necessary:
1) to complement the information on the neutral IGM component derived from the
Lyman forest, checking the associated metal absorption lines (in particular due to C IV and Si IV) and deriving the contribution of the ionised gas; 2) to complete the coverage of the associated HI and metal transitions. For this purpose spectra of the same targets have to be obtained (e.g. at higher resolution with UVES/VLT - simultaneously via a fiber link) or retrieved from the archives.

An exposure time of $\sim 13$ hours would be needed with CUBES to observe a sample of $\sim 40$ bright background quasars at $z_{em} \sim 2-3$, reaching the necessary signal-to-noise ratio\cite{Dodorico2022}. The improved performances of CUBES with respect to UVES/VLT will open up background sources 1-2 magnitudes fainter than the quasars used at present, significantly increasing the number (and the space density) of available targets and/or the possibility to obtain very high signal-to-noise ratio spectra to detect the faint lines.

\subsection{Cosmic UV background}
Galaxies are thought to be able to produce the bulk of the UV emissivity need for cosmic reionisation at high redshift, but quasars probably also contribute, with other more esoteric possibilities also proposed (e.g. decaying particles). Estimates of the escape fraction (\fesc) of hydrogen-ionising photons able to escape a galaxy are close to 100\% for quasars. However, the volume density of low- and intermediate-luminosity quasars at $z$ $>$ 4 is still uncertain, so it is unclear if they are the dominant source of ionisation. In contrast, star-forming galaxies are more numerous, but estimates of \fesc from observations of the Lyman continuum ($\lambda_{\rm rest}$ $<$ 91.2 nm) have uncertainties of tens of percent and are limited to a handful of systems at $z$ = 2.5 to 4. The \fesc value for each galaxy depends on factors such as star-formation intensity, feedback from supernovae, the depth of galactic potential wells, and the dynamics of the absorbing gas. To be detected by an observer (at $z$ = 0), escaping photons have to survive absorption along the line of sight by the intergalactic medium, which increases strongly with redshift and is significantly variable between sightlines. Given these competing factors, the ideal redshift range for ground-based observations of the Lyman continuum of a galaxy is $z$ = 2.4 to 3.5, i.e. from about 410 nm down to the atmospheric cut-off. In short, CUBES is essential to help progress this topic. Its sensitivity will enable at least an order of magnitude increase in the number of systems with estimates. Moreover, with unrivalled sensitivity at the shortest wavelengths, it will provide strong constraints on \fesc at redshifts of $z \sim 2.5$, where the line of sight absorption is much reduced compared to higher redshift systems.

\subsection{Transient astronomy}
Time-domain astronomy is one of the most active branches of modern astrophysics. In a few years, new observational facilities, specifically designed with the goal of securing high-cadence observations of large fractions of the nightly sky, will become operational. Equally important, ``big-data'' algorithms, in a full machine-learning scenario, are increasingly being applied and developed to manage the large amount of data provided by these facilities. All in all, we are facing an impressive technological and conceptual effort that will allow the research community to identify thousands-to-millions of new or recurrent transient sources routinely. The discovery space opened by rare or peculiar transients is highly relevant, involving all categories of sources, from Galactic novae and unusual variables to supernovae, GRBs, and gravitational-wave electromagnetic counterparts, not to mention possible entirely new classes of objects. For low or high redshift objects, a highly efficient UV spectrograph can shed light on a variety of physical ingredients: transient chemical composition, ejecta dynamics, accretion profiles, early-time kilonova evolution, time-resolved line profile for, e.g., GRBs and other “explosive” events, etc. In this context, the possible synergy of the CUBES and UVES spectrographs could open the exciting perspective of a UV-blue continuous spectral coverage, allowing the ESO community to access higher redshift sources and a substantially larger set of diagnostic spectral lines.
\subsection{A treasure trove of science}
Given its efficiency, the spectral range and the resolution, CUBES will not only be able to fulfill the main scientific objectives, but also open a new parameter space in observational astronomy \cite{Alcala2022, Ali2022, Balashev2022} with hopefully many new and unexpected results. The current workhorse for quantitative spectroscopy at these short wavelengths is UVES, but its efficiency drops to just a few percent in the near-UV. This means that in a broad range of scientific topics  we are limited to studies with relatively small samples of objects. The driving philosophy for CUBES is to provide a significant efficiency gain and open-up an exciting new capability for ESO’s Paranal Observatory.
\section{FROM SCIENCE TO REQUIREMENTS}
\label{sec:requirements}  
The science cases of interest for the CUBES community have been used to flow-down the Top level Requirements (TLR) in Phase A and to effectively contribute in the design trade-offs, via use of software tools developed in the study (ETC, E2E simulator), both in Phase A and in the current Phase B. 
Key TLRs, identiﬁed for the development of the instrument conceptual architecture and design, were:
\begin{itemize}
\item Spectral range: CUBES shall provide a spectrum of the target over the entire wavelength range of 305 – 400 nm in a single exposure (goal: 300 – 420 nm).
\item Eﬃciency: The eﬃciency of the spectrograph, from slit to detector (included), shall be $>40$\% for 305 – 360 nm (goal $>45$\%, with $>50$\% at 313 nm), and $>37$\% (goal $40$\%) between 360 and 400 nm.
\item Resolving power ($R$): In any part of the spectrum, $R$ shall be $>19000$, with an average value $>20K$, where R is deﬁned as the full width at half maximum (FWHM) of unresolved spectral lines of a hollow cathode lamp in the spectral slice.
\item Signal-to-noise (S/N) ratio: In a 1 hr exposure the spectrograph shall be able to obtain, for an A0-type star of $U$ = 17.5 mag (goal $U \geq 18$ mag), a S/N = 20 at 313 $nm$ for a 0.007 $nm$ wavelength pixel. For different pixel sizes, the S/N ratio shall scale accordingly.
\end{itemize}

An important development in the Phase A study was the potential provision of a second (lower) resolving power (with $R \sim 7$k), to enable background-limited observations of faint sources where spectral resolution is less critical. 
We have also investigated a potential fiber-feed to UVES, to provide simultaneous observations at longer wavelengths. This broadens the scientific capabilities of CUBES (by significantly enhancing the cases related to e.g. transients) while also offering operational efficiencies for many cases where observations at longer wavelengths are required to support the UV analysis.

\begin{figure}
    \centering
    \includegraphics[width=1.0\textwidth]{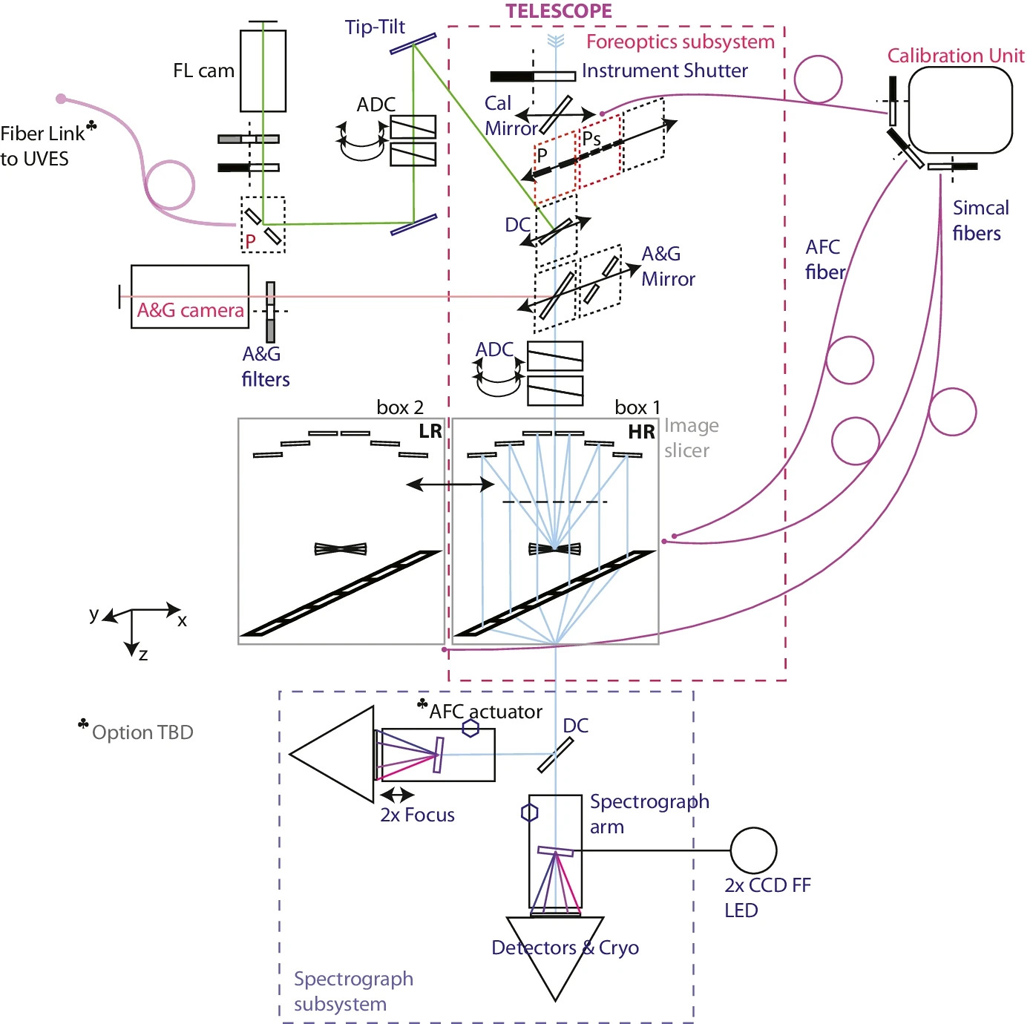}
     \caption{Functional scheme of the CUBES system (in which the light path goes from top to the bottom). The following acronyms are used: DC - dichroic, P - pinhole, Ps - series of pinholes, A\&G - Acquisition and Guiding, AFC - Active Flexure Compensation system, FL - fiber link, ADC - Atmospheric Dispersion Corrector. Optional modules (TBD in Phase-B) are marked with a trefoil sign \cite{Zanutta2022}. }
    \label{fig:FunctionalScheme}
\end{figure}

The conceptual optical architecture and functional scheme derived from the Phase A study is shown in  Fig.\ref{fig:FunctionalScheme}. 
At system level, the identification of instrument active functions (sliding mechanisms, ADCs, filter wheels, tip-tilt mirrors, refocusing mechanisms) drove the definition of requirements in term of mass to be moved, motion range and speed, repeatability.
After the definition of the technical specifications, their flow down to the different work-packages/subsystems along with the proper management of requirement justification has been carried out through the adoption of Model-Based System Engineering approach and related tools (e.g.~Cameo and SysML \cite{Casse2017}). As described in \cite{Zanutta2022} , in order to cope with different expertise and the time-frame of the project, the modeling environment has been restricted to the system-engineering team while interacting with the other engineers using traditional Excel ﬁles. Other relevant system-level activities started in the end of Phase A and are currently on-going in Phase B: the setting and management of system budgets interconnected to the requirements (physical, operational and performance) flow-down to subsystems and the definition of subsystems interface requirements.
\section{INSTRUMENT DESIGN OVERVIEW}
\label{sec:design}  
The instrument has two modes, either using CUBES alone or combined with the fiber-feed to UVES. Two spectral configurations ($R > 20K$ and $\sim 7$) are enabled by the exchange of two independent image slicers. An Active Flexure Compensation system (AFC) is not included in the proposed baseline, but is an option being considered if required.
All the top level requirements (TLRs), in particular those related to efficiency, are met.
For the use of CUBES we have identified no significant operational issue deviating from common practice with other instruments at Paranal.

\subsection{Instrument sub-systems and Operations}
In order to achieve its scientific goals the current baseline design of CUBES includes:
\begin{itemize}
\item Calibration unit that provides the light sources necessary to register frames for ﬂat ﬁelding, wavelength calibration, alignment, the options of simultaneous wavelength calibration and AFC if required.
\item	A foreoptics sub-system that includes an atmospheric dispersion corrector (ADC) and an Acquisition and Guiding functionalities.
\item Two image slicers (to enable different spectral resolutions).
\item A spectrograph subsystem, composed of two arms, both equipped with transmission gratings, with a high groove density and working at first order, and cameras. Each arm has its own detector cryostat, which comprises a 9k or 10k CCD as detector, read-out electronics, cryo-vacuum components (both hardware and specific control electronics).
\item Instrument Control Electronics, based on PLC compliant to latest ELT design electrical standard, which has the role to control all the functions in the instrument,  excluding  the  Scientific Detector system and its associated Cryostat and Vacuum controller.
\item Instrument Software comprising by control software, data-reduction and simulation tools (see section \ref{sec:Software}).
\item A fiber-Link unit provides the option of simultaneous observations with UVES\cite{UVES2000} (in its red band 420-1100 nm), via relay optics feeding optical fibers (1 object, 3 sky) subtending 1 arcsec aperture and approximatively 40 m long to transmit light from UT2 Cassegrain to Nasmyth platform.
\end{itemize}
\begin{figure}[ht]
    \centering
    \includegraphics[width=0.9\textwidth]{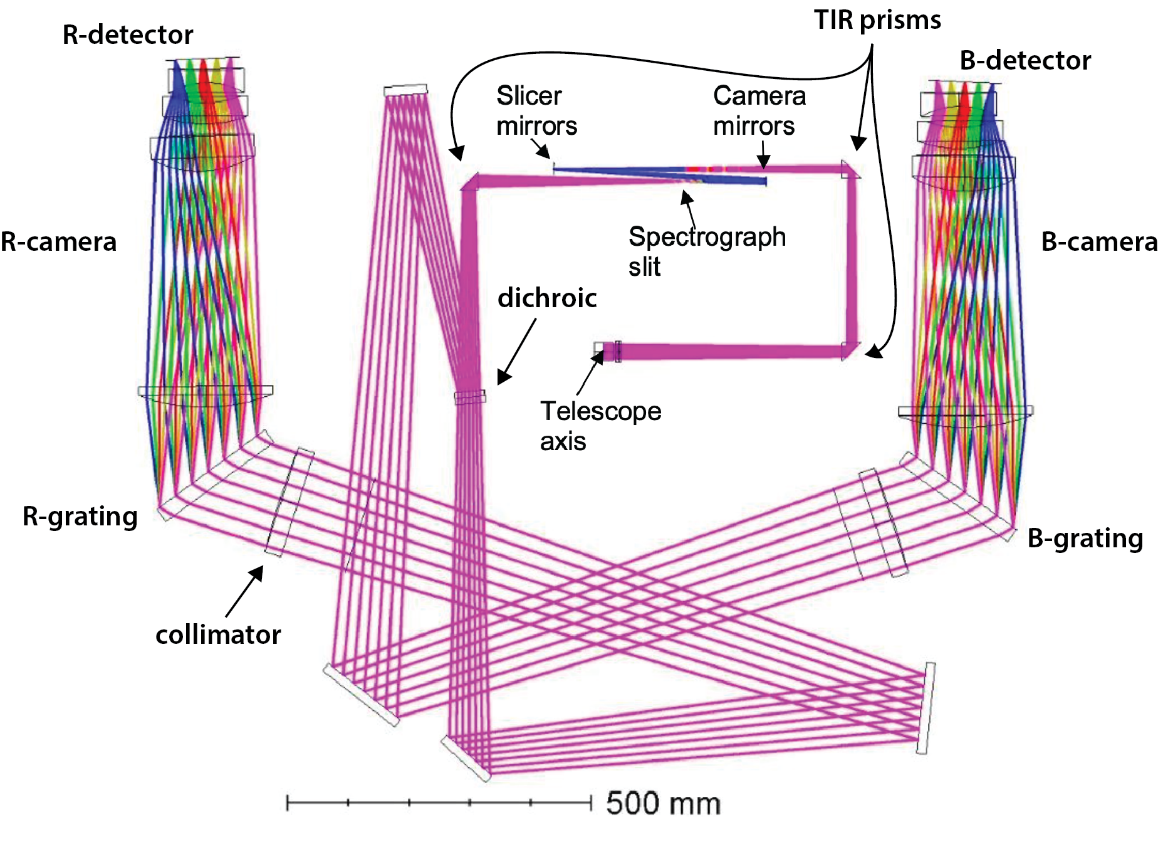}
     \caption{Optical end-to-end model of the CUBES instrument.}
    \label{fig:CUBES_optic}
\end{figure}
\subsection{Optics}
Using two lens doublets and a number of folding prisms (not shown in Fig. \ref{fig:FunctionalScheme}), the foreoptics relays a FoV of 6”x10” at the telescope focus to the entrance plane of the spectrograph. Direct-vision ADC prisms in the parallel beam between the doublets provide atmospheric dispersion correction over a range 300-405 nm for zenith angles of 0-60°. By inserting a dichroic just below the telescope focal plane, light redward of 420 nm may be directed to the UVES fiber feed. During acquisition the object field is directed by a 45 mirror to the A\&G CCD which is equipped with a set of photometric filters. After acquisition the mirror is moved to pass the center of the field to the spectrograph.
At the magnified telescope focal plane produced by the foreoptics (scale 0.5 mm/arcsec), one of two user-selectable reflective image slicers decomposes the rectangular FoV into six slices. Six camera mirrors, one for each slice, re-image the slices on an output slit mask. Through optimised dielectric coatings and careful mask alignment, the slicer efficiency is expected to be $>90\%$ (goal 94\%) The output slit mask has six slitlets, corresponding to six slices, each one measuring 0.25”x10” on the sky for the HR slicer ($R \sim 20K$) and 1”x10” for the LR slicer ($R \sim 7K$). Further slit mask apertures are illuminated by a ThAr fiber source for simultaneous calibration and/or use by the AFC system.

The light coming from the slit mask is folded by a TIR prism and then reaches a dichroic which splits the light by reflecting the Blue-Arm passband (300–352.3 nm) and transmitting the Red-Arm passband (346.3–405 nm), exceeding the 305-400 nm TLR. The layout of the two arms is similar but the individual components and separations are different so as to achieve the required dispersion, magnification and image quality for the 2 passbands, using only fused silica. The F/20 collimator is a single lens. The spectrograph camera is composed of 3 aspheric, tilted and decentered lenses.

Also the large CCDs (9K$\times$9K or 10K$\times$10K) are tilted to compensate for the change of refractive index of Silica with wavelength. The CCDs are cooled with Stirling coolers and controlled by ESO’s 2nd-generation detector control system (NGC2).

\begin{figure}[ht]
    \centering
    \includegraphics[width=0.7\textwidth]{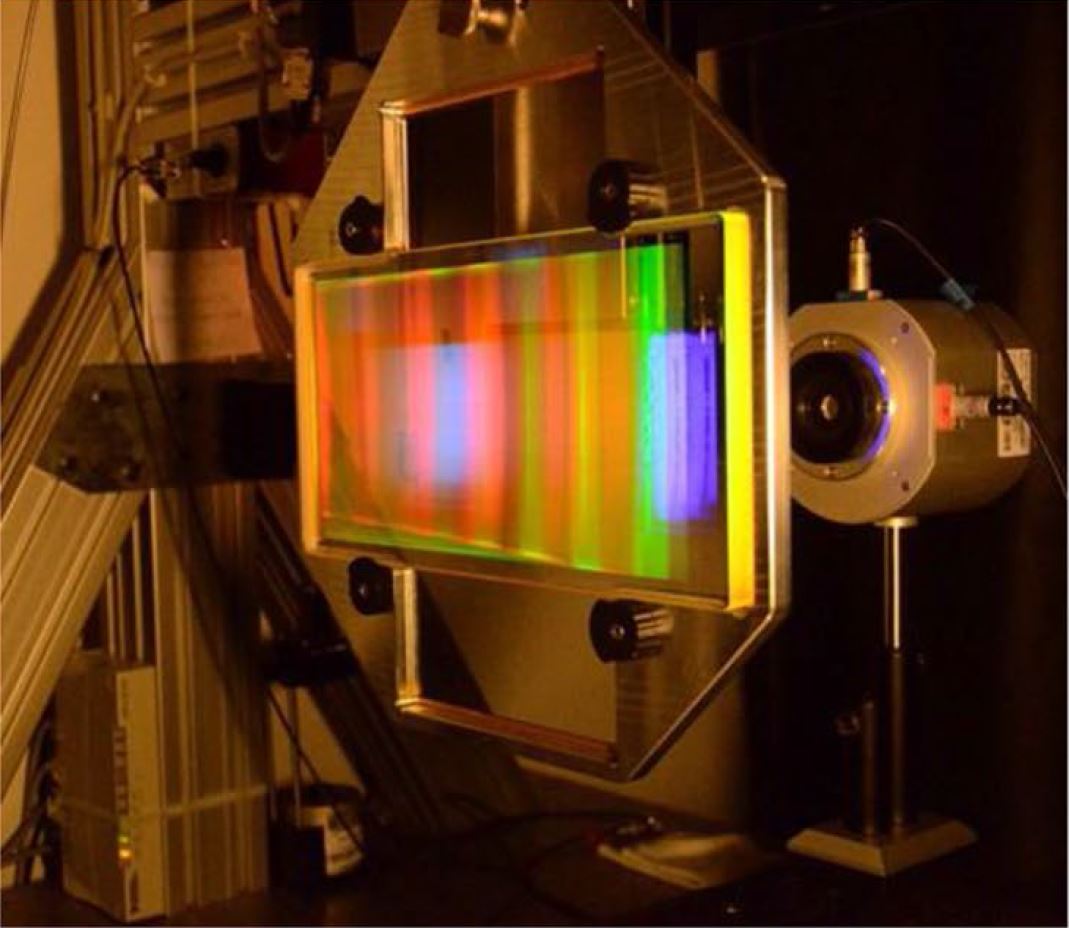}
     \caption{A CUBES grating prototype mounted in the setup available at the Fraunhofer Institute for Applied Optics and Precision Engineering, Jena, for the measurement of the diffraction efficiency.}
    \label{fig:CUBES_grating}
\end{figure}
\begin{figure}
    \centering
    \includegraphics[width=0.8\textwidth]{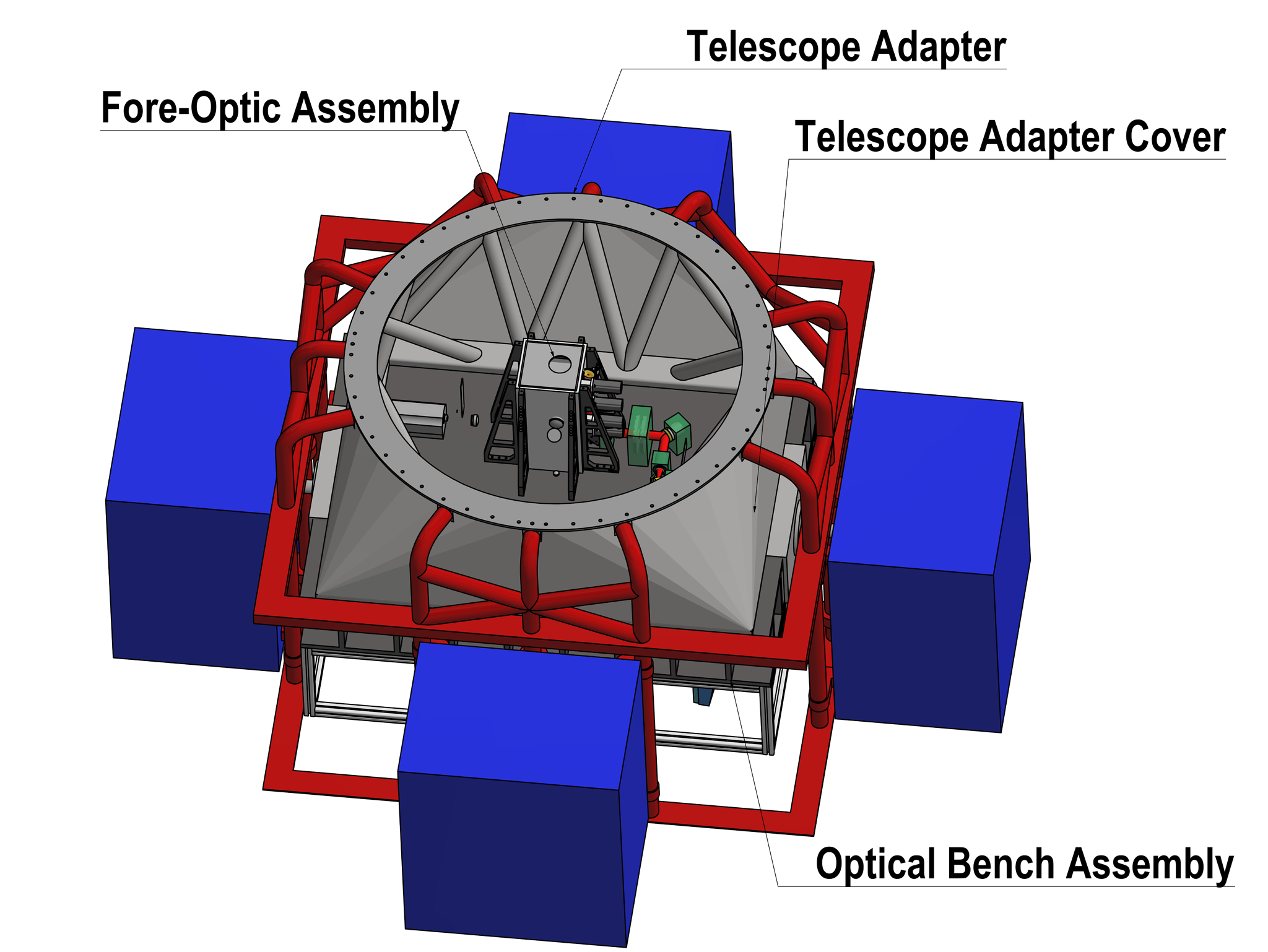}
     \includegraphics[width=0.8\textwidth]{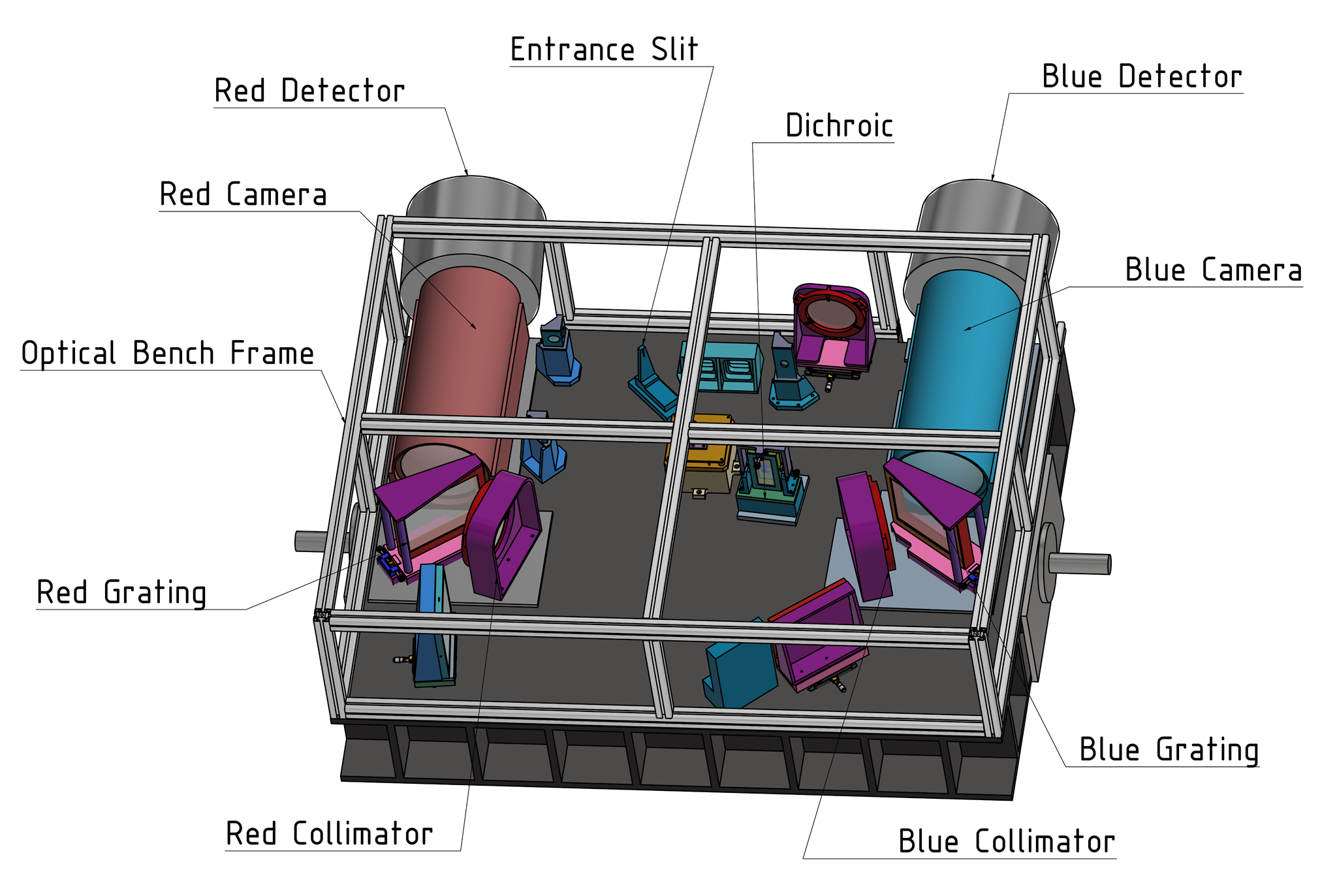}
    \caption{Mechanical concept for CUBES: The general layout of the main mechanical components is shown in the upper image.
For reference, the Telescope Adapter has a diameter of about 1.5 m. The Fore-Optics Assembly is located near the center of
the bench, while the spectrograph opto-mechanics are `hanging' on the bottom side of the same bench (lower image)
    \cite{Zanutta2022}.}
    \label{fig:MechanicalConcept}
\end{figure}
\subsubsection{The Dispersing Elements}
In order to achieve a high ($>20K$) resolution without the efficiency losses associated with crossdispersed echelles, CUBES uses state-of-the-art first-order dispersing elements. Binary transmission gratings produced by E-beam microlithography and an ALD overcoat have been identified as a suitable technology (see \cite{Zeitner2022} and the paper by Zeitner et al. at this conference). Their theoretical average (RCWA) diffraction efficiency is $> 90\%$. A first prototype, funded by FAPESP, was produced and tested as early as 2018, further prototyping activity is underway and a test and characterization report will be presented at the PDR. 
\begin{figure}[hb]
    \centering
    \includegraphics[width=0.525\textwidth]{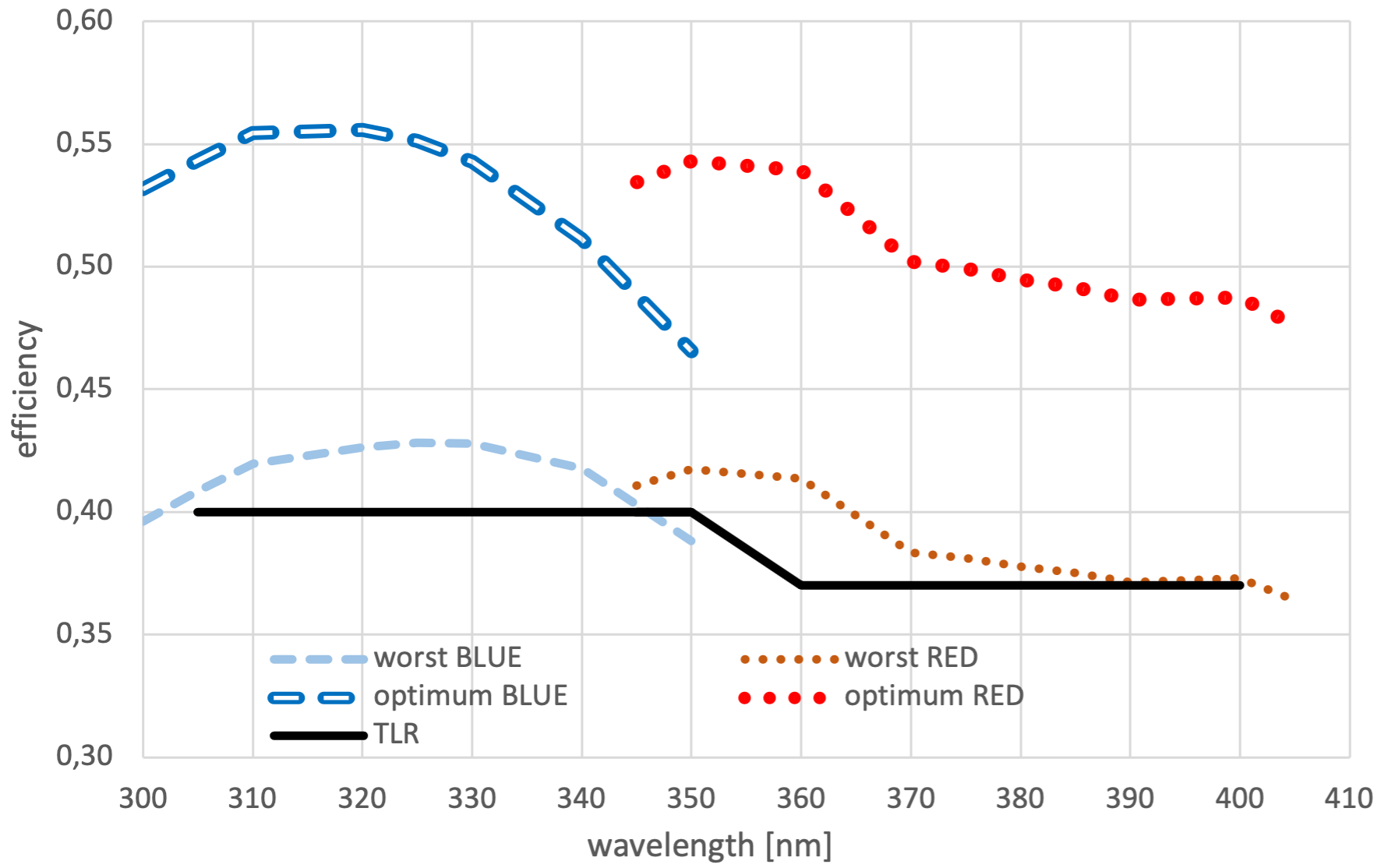}
        \includegraphics[width=0.465\textwidth]{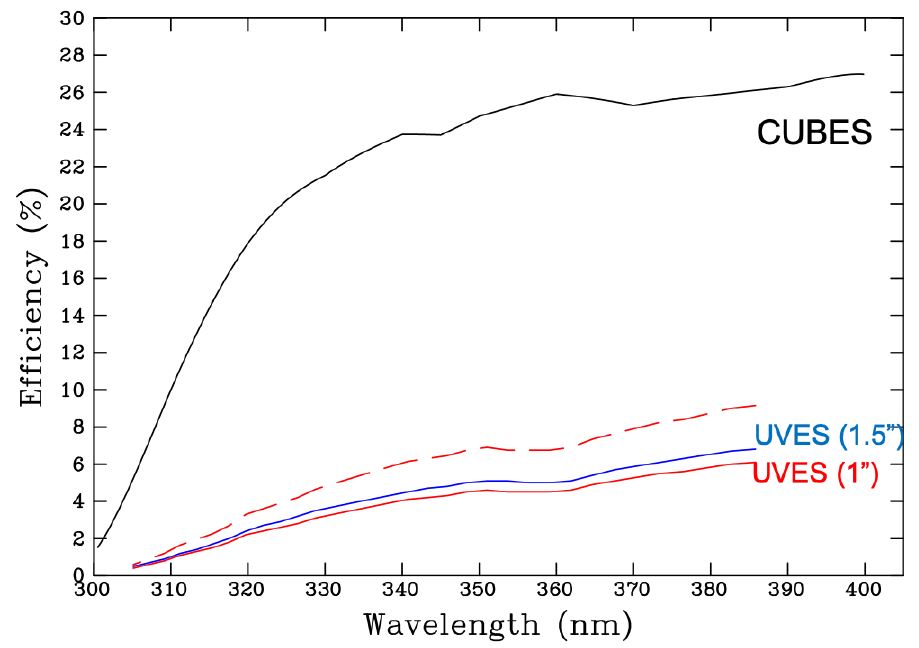}
     \caption{{\it Left}: Calculated detective quantum efficiency of the two arms of CUBES (BLUE and RED) for worst and best scenarios. The black line indicates the formal top-level efficiency requirement. The throughput was calculated from the telescope focus, including detector QE, IS vignetting, but excluding atmospheric effects and telescope losses. {\it Right}: Comparison of predicted CUBES efficiency (including telescope and atmosphere) with those for the central wavelengths of the UVES echelle orders from the ESO ETC. The dashed red line shows the anticipated gain in performance ($\times 1.5$) from a possible UVES upgrade.
    \label{fig:efficiency}
    }
\end{figure}
\subsection{Mechanics}
CUBES requires a fairly large beam diameter of 160 mm. Consequently, the instrument envelope is also rather large  compared  to  other  Cassegrain  instruments  (cf. X-Shooter with  a  100 mm  beam  diameter).  Scaling  classical  instrument  designs  to  the  required  size of  CUBES  would  result  in  exceeding  the  mass  limit of 2500 kg for UT (Unit Telescope) Cassegrain instruments. Therefore, a light-weight construction principle has been adopted, making use of modern composite materials. The optical layout was optimized such that all optical elements of the spectrograph from slit to detector lie in a single plane, so all spectrograph optics can be mounted on a single optical bench with size 1.3×1.7 m. This is arguably the most stable configuration since the dispersion direction of  CUBES  is  parallel  to  the  stiff  surface  plane  of  the optical bench. A general focus of the mechanical design is, in fact, to minimize the effects of gravitational bending of the instrument.

In the current design, the CUBES main mechanical structure is divided into three main components: 1.  A telescope adapter, that provides a stiff connection between the Cassegrain telescope flange and the optical bench assembly; 2.  An optical bench that provides a stable platform for the spectrograph optics as well as for the foreoptics; 3.  An assembly to provide support for auxiliary  equipment  such  as  electronic racks, the calibration unit and vacuum equipment. This frame is detached from the optical bench to mitigate the contribution of flexure. We are currently planning to use steel for the telescope adapter and the support frame, and a Carbon-fiber reinforced polymer (CFRP) for the optical  bench.  The  optomechanics  are  currently  considered  to  be  made  of  aluminum  alloys,  e.g. AlSi40, to  improve  the  specific  stiffness  and  lower  the Coefficient of Thermal Expansion (CTE) mismatch between the optomechanical parts and the CFRP  bench.

\subsection{Software}
\label{sec:Software}
\begin{figure}
    \centering
    \includegraphics[width=1.0\textwidth]{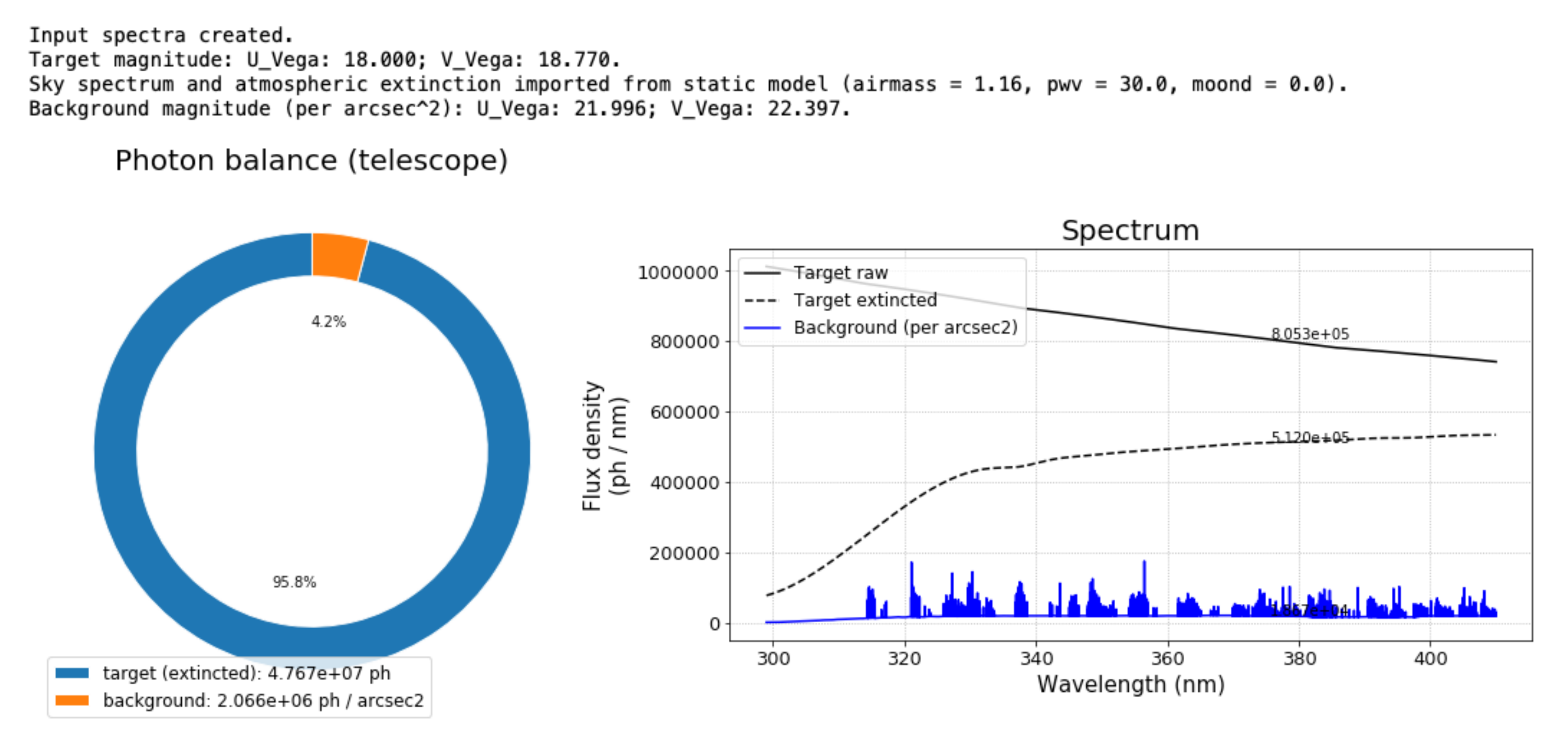}
     \caption{Example of the input spectra in the \textit{Basic Version} E2E notebook (for a flat spectrum of $U$\,$=$\,18\,mag). The integrated flux from the target and the sky is computed assuming the collecting area of the primary mirror of a VLT unit telescope and a detector integration time of 3600 s. \textit{Left:} photon balance between the target and the sky background. \textit{Right:} spectra of target (with extinction) and background.
    \label{fig:spectrum simulator}
    }
\end{figure}

\begin{figure}
    \centering
    \includegraphics[width=0.43\textwidth]{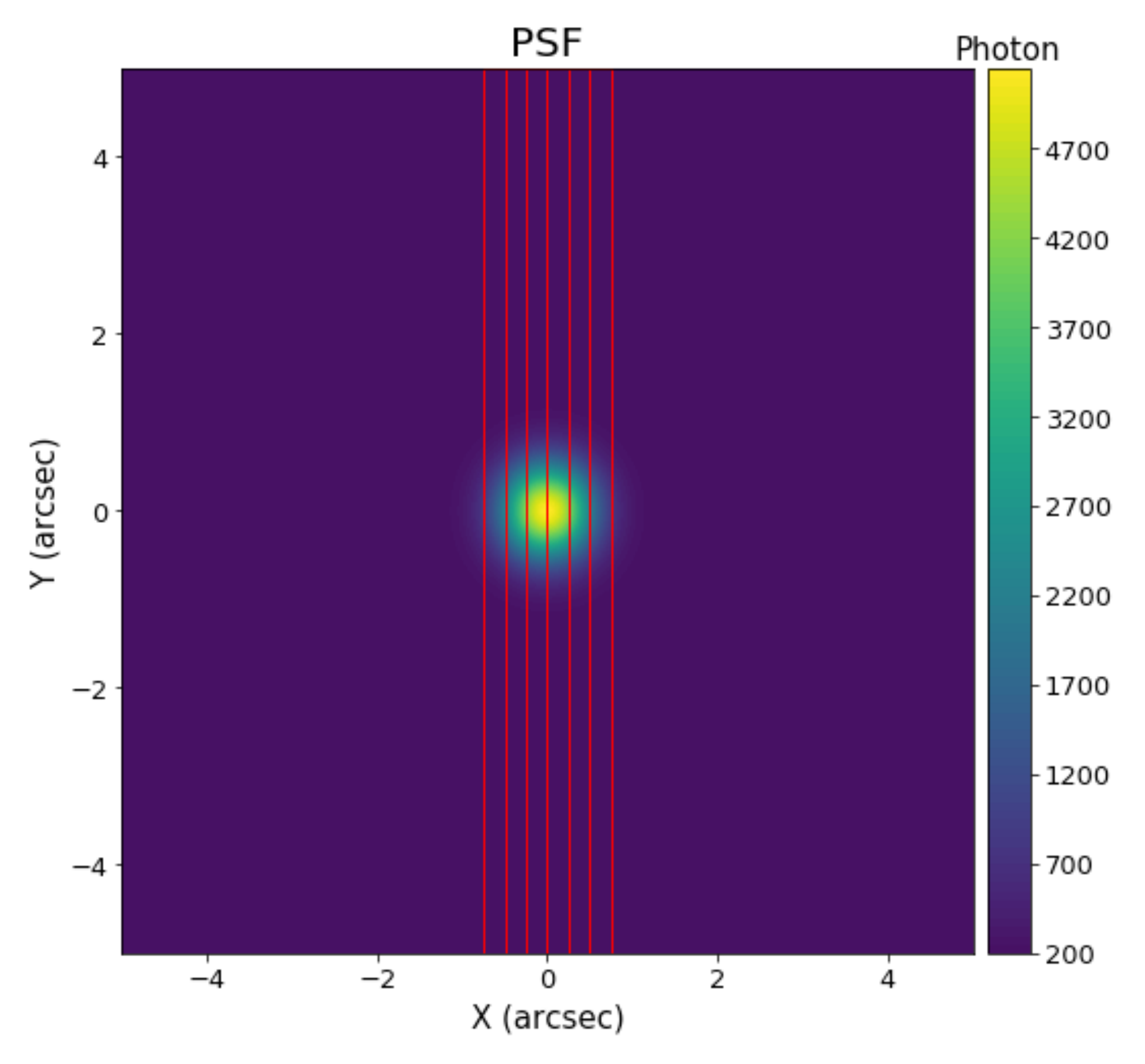}
    \includegraphics[width=0.56\textwidth]{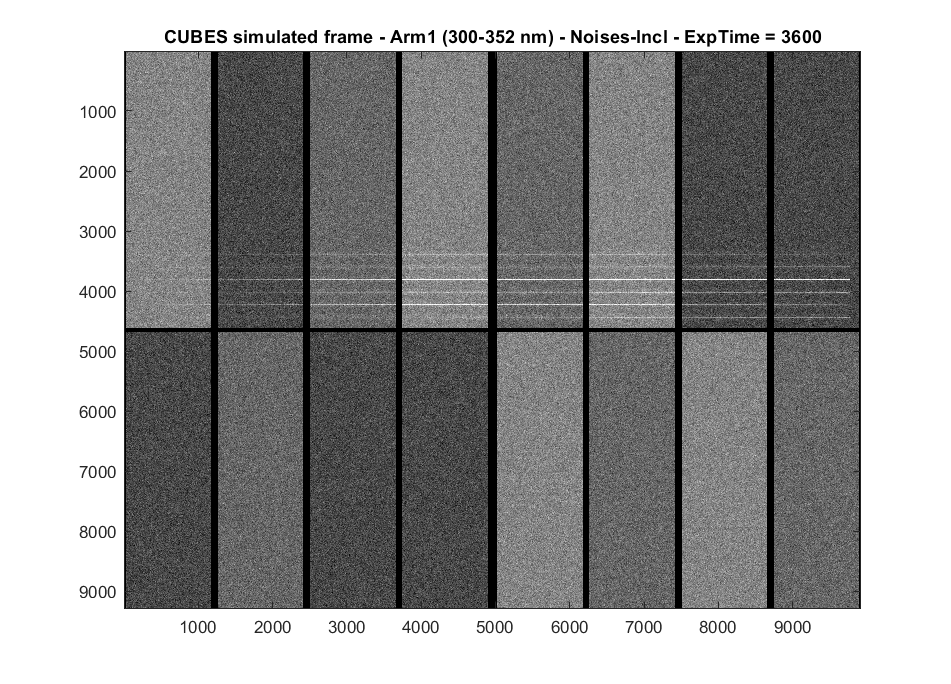}
    \caption{{\it Left}: Simulated image of the target point spread function (PSF) on the (high-resolution) slicer focal plane, with the slice boundaries superimposed in red. {\it Right:} Simulated raw frame for Arm 1. The six slice traces are projected onto the upper half of the detector to allow the lower half to be read out separately to record the active flexure compensation (AFC) system (if analysis in the next phases shows it is required).
    \cite{Genoni2022}}
    \label{fig:psf_slices_frame}
\end{figure}

The CUBES instrument benefits from an entire  {\it ``software ecosystem''} built around it, whose individual packages cooperate to support the users from the proposal preparation to the science-grade spectra:
\begin{itemize}
\item The Exposure Time Calculator (ETC): a web application used to predict  the CUBES exposure time required to achieve a given SNR (see Fig.~\ref{fig:CUBES_SNR_mag});
\begin{figure}[ht]
    \centering
    \includegraphics[width=0.75\textwidth]{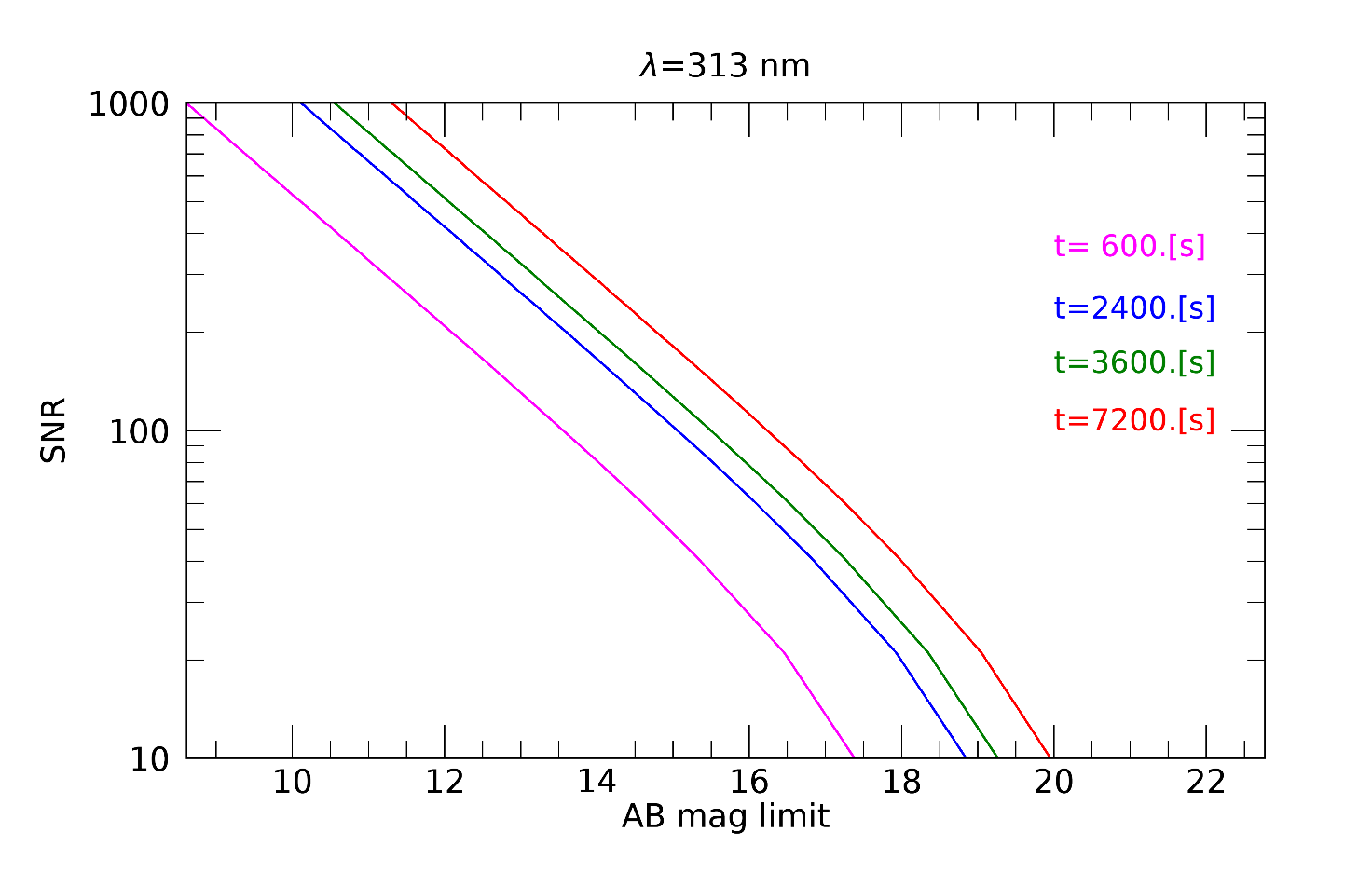}
     \caption{SNR versus magnitude limit (AB unit) at 313nm for different exposure times, in HR-mode, template spectrum with constant AB magnitude, $U_{sky}$ = 22 mag arcsec$^{-2}$, airmass= 1.16, seeing = 0.87", dark current= 0.5e- pix$^{-1}$ hr$^{-1}$, RON = 2.5 e- rms, and binning 1x1 pixel.}
    \label{fig:CUBES_SNR_mag}
\end{figure}
\item The Observation Preparation Software (OPS): a list of tools aimed to help the users identify the best instrument settings to achieve a scientific goal;
\item The Instrument Control Software (ICS): devoted to the instrument devices and detectors control, as well as to the execution of the high level procedures aimed to coordinate the science exposures and produce the final files to be ingested in the archive;
\item The Data Reduction Software (DRS): a collection of C++ and Python recipes aimed to remove instrument artifacts from the science exposures, and produce the final 1D spectrum;
\item The End-to-end Simulator (E2E): a package able to simulate realistic science exposures of a given science target, by taking into account the CUBES instrumental effects, and allowing early testing of the data reduction pipelines, as well as early validation of design decisions (Figs.~\ref{fig:spectrum simulator} and \ref{fig:psf_slices_frame}).
\end{itemize}
A relevant feature of many of the above mentioned packages is that they will be developed according to the recently published ELT software standards, and will be based on the ELT Instrument Framework libraries and tools.  The CUBES instrument will, however, operate in a VLT environment (the Paranal Observatory) hence a special component is being developed by ESO to allow proper communication between the two environments.  Such component, named VLT/ELT Gateway, will also be used to communicate with the UVES instrument and coordinate the CUBES+UVES simultaneous exposures from the CUBES control software. The CUBES software ecosystem is described in detail in \citenum{2022-Calderone_SPIE}.

\section{PROJECT and MANAGEMENT}
\label{sec:management}  
\begin{figure}
    \centering
    \includegraphics[width=1.0\textwidth]{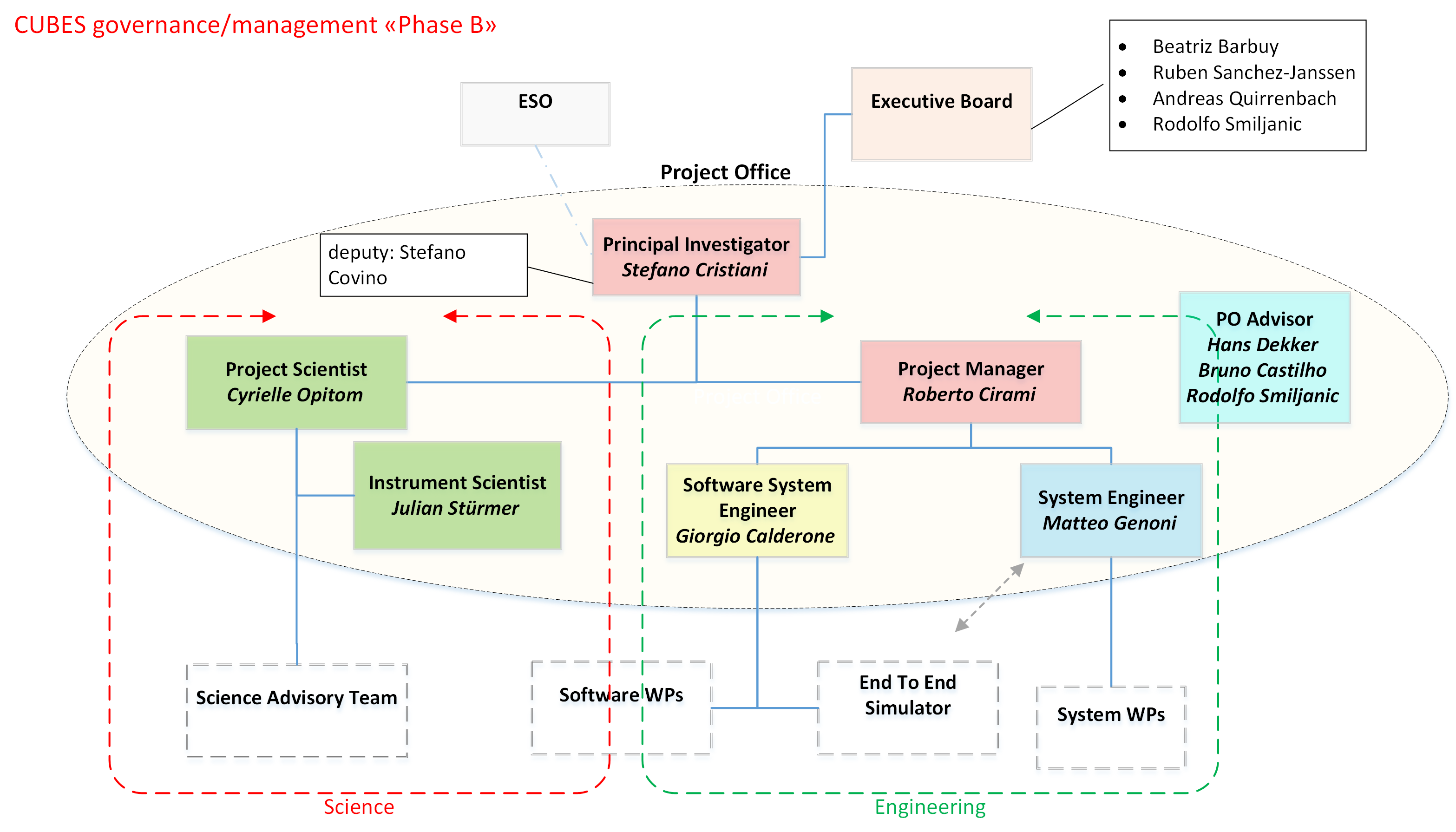}
     \caption{Consortium organizational chart.}
    \label{fig:CUBES_org_chart}
\end{figure}
The CUBES consortium is composed of institutes from five countries:
\begin{itemize}
\item	INAF - Istituto Nazionale di Astrofisica, Italy, (consortium leader)
\item IAG-USP - Instituto de Astronomia, Geofísica e Ciências Atmosféricas (primary Brazil partner) and LNA - Laboratório Nacional de Astrofísica (secondary Brazil partner), Brazil
\item LSW - Landessternwarte, Zentrum für Astronomie der Universtität Heidelberg, Germany
\item	NCAC - Nicolaus Copernicus Astronomical Center of the Polish Academy of Sciences, Poland
\item	STFC-UKATC - UK Astronomy Technology Centre, (primary UK partner) and Durham University Centre for Advanced Instrumentation (secondary UK partner), United Kingdom 
\end{itemize}

The Consortium is organized in a rather standard way, with a Principal Investigator (PI) who has the ultimate responsibility of the project and is the formal contact point between ESO and the Consortium. The PI represents the leading technical institute, INAF. Each country is represented in the managerial structure by one Co-PI; altogether form the CUBES Executive Board (EB). The EB is a strong but hands-off high-level overview committee. The managerial aspects are delegated by the EB to the Project Manager (PM). The scientific aspects are delegated by the EB to the Project Scientist (PS). The project manager is supported by a System Engineer (SE) and by a Software System Engineer (SSE) who are in charge to supervise the overall system design. The SE and SSE work in close contact with the Instrument Scientist (IS) who makes sure that the adopted technical solutions match the foreseen instrument scientific needs. 
The overall work is organized in work-packages coordinated by the CUBES system engineers. There is one responsible per work package, named Work Package Manager (WPM) who is in charge to organize tasks, resource allocation, interfaces, dependencies and budgeting of the respective work package. The PM integrates information provided by each WPM, reviewed by SE and SSE, to obtain a complete overview of the project in terms of cost, manpower and work performed keeping track of the overall project status.
\begin{figure}
    \centering
    \includegraphics[width=1.0\textwidth]{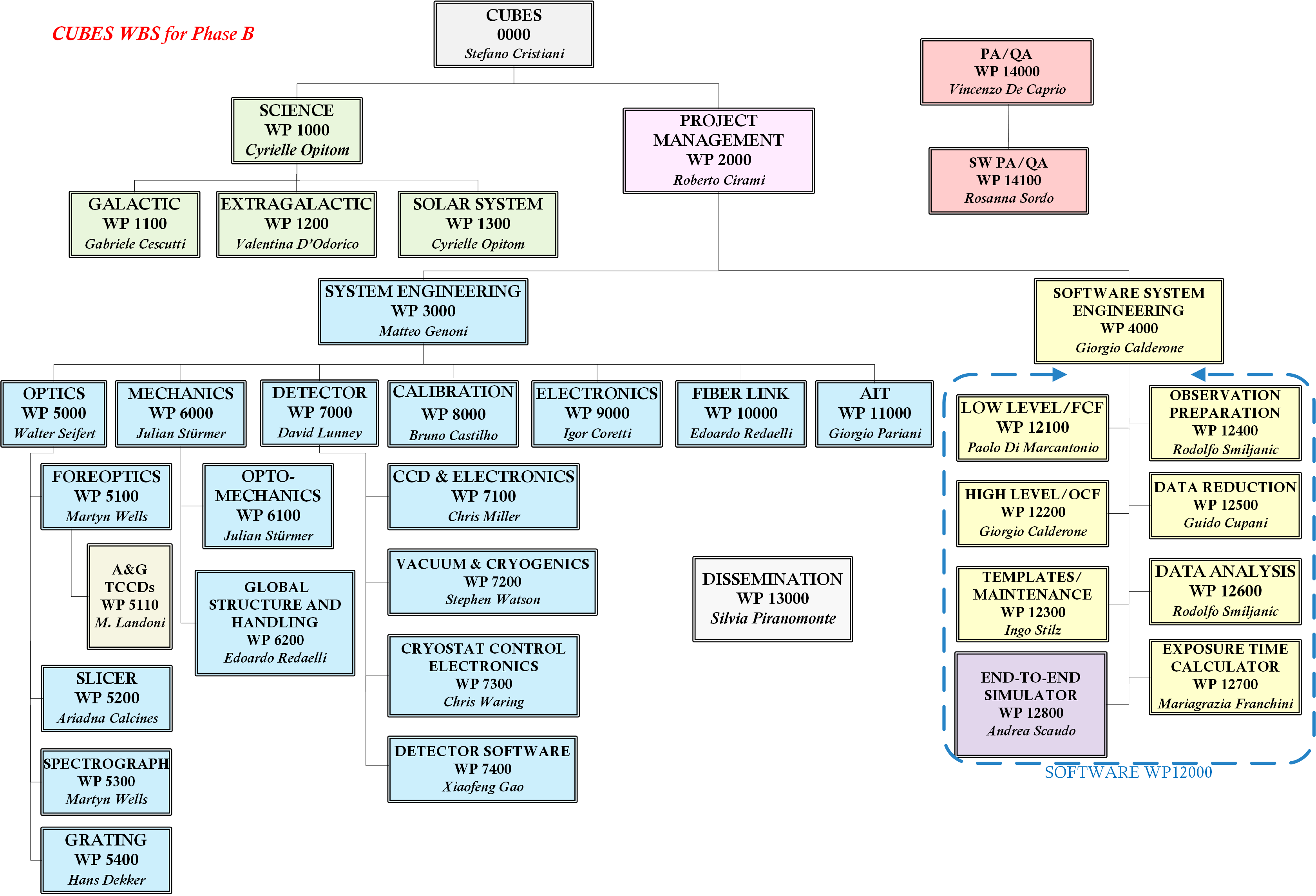}
     \caption{CUBES Work Breakdown Structure.}
    \label{fig:CUBES_WBS}
\end{figure}
\begin{table}
\caption{CUBES organisation breakdown structure.} 
\label{tab:CUBES_organization}
\begin{center}       
\begin{tabular}
{|p{0.30\linewidth} | p{0.26\linewidth} | p{0.22\linewidth}|}
\hline
\rule[-1ex]{0pt}{3.5ex}  {\bf Role} & {\bf Name} & {\bf Institute}\\
\hline
\rule[-1ex]{0pt}{3.5ex}  Principal Investigator & Stefano Cristiani & INAF-OATs\\
\hline
\rule[-1ex]{0pt}{3.5ex}  Deputy-PI              & Stefano Covino    & INAF-OABr \\
\hline
\rule[-1ex]{0pt}{3.5ex}  Co-PI & Beatriz Barbuy & IAG-USP  \\
\hline
\rule[-1ex]{0pt}{3.5ex}  Co-PI & Ruben Sanchez-Janssen & STFC-UKATC \\
\hline
\rule[-1ex]{0pt}{3.5ex}  Co-PI & Andreas Quirrenbach & LSW\\
\hline 
\rule[-1ex]{0pt}{3.5ex}  Co-PI & Rodolfo Smiljanic & NCAC\\
\hline 
\rule[-1ex]{0pt}{3.5ex}  Project Scientist & Cyrielle Opitom & University of Edinburgh\\
\hline 
\rule[-1ex]{0pt}{3.5ex}  Project Manager& Roberto Cirami & INAF-OATs  \\
\hline 
\rule[-1ex]{0pt}{3.5ex}  Instrument Scientist& Julian Stürmer& LSW  \\
\hline 
\rule[-1ex]{0pt}{3.5ex}  System Engineer& Matteo Genoni & INAF-OABr  \\
\hline 
\rule[-1ex]{0pt}{3.5ex}  Software System Engineer& Giorgio Calderone &  INAF-OATs  \\
\hline 
\end{tabular}
\end{center}
\end{table}
\subsection{Project phasing and Schedule}
CUBES adopts the standard project phasing for ESO instruments which is based on the stage-gate paradigm. Important decision points are project milestones (gates of the project) which mark the transition into a new stage when successfully completed. 
The Phase A study commenced in June 2020, with a close-out review in June 2021. The next phase of the project started in February 2022 with $\sim 2.5$ yrs for the detailed design, with two formal review milestones of the Preliminary and Final Design Reviews. The project will then enter the Manufacturing, Assembly, Integration, Testing (MAIT) phase, with the next major milestone being the system Integration and Test Readiness Review (ITRR), which will be closed by the Preliminary Acceptance Europe
(PAE). CUBES will then be moved to Chile, reassembled,
mounted on the VLT and tested. The commissioning
phase will be closed when the Provisional
Acceptance Chile (PAC) is granted. Finally, CUBES
will be offered to the community after PAC is granted
and science verification is concluded. The MAIT phase
through to science operations in the current plan is envisaged
for 3 yrs, meaning that it would be available to
the ESO user community in 2028.
\begin{table}[ht]
\caption{Project plan per phases.} 
\label{tab:ProjectPlanPhases}
\begin{center}       
\begin{tabular}
{|p{0.23\linewidth} | p{0.2\linewidth} | p{0.35\linewidth}|}
\hline
\rule[-1ex]{0pt}{3.5ex}  {\bf Project Phase} & {\bf Duration} & {\bf Remarks}\\
\hline
\rule[-1ex]{0pt}{3.5ex}  Entry into force of the Construction Agreement & T0 & 2022/Feb/15\\
\hline
\rule[-1ex]{0pt}{3.5ex}  Phase B & T0 + 10 months & Preliminary design with review concluded \\
\hline
\rule[-1ex]{0pt}{3.5ex}  LLI - Optical FDR & B+7 months & Optics finalization  \\
\hline
\rule[-1ex]{0pt}{3.5ex}  Phase C & B + 16 months & Final design with review concluded and LLI orders placed  \\
\hline
\rule[-1ex]{0pt}{3.5ex}  Phase D & C + 36 months & MAIT with PAE\\
\hline 
\rule[-1ex]{0pt}{3.5ex}  Phase E - PAC & D + 14 months & AIV in Chile, commissioning concluded and PAC\\
\hline 
\rule[-1ex]{0pt}{3.5ex}  {\bf Total:} & 77 months & Construction, Commissioning and PAC  \\
\hline 
\end{tabular}
\end{center}
\end{table}

\subsection{Public Engagement}
CUBES is an ambitious research program, and some of the scientific topics are related to the hottest open questions in modern astrophysics. Considering the vast discovery potential of the project, and the remarkable research and development technological activities, we consider good public communication as particularly important. Dissemination of science and technology is a fundamental part of our project and  since the beginning we have included a work package devoted to public outreach. During phase A we have mainly communicated the progress of our project and the main scientific topics. We have prepared a web page that is also a useful tool for the project as a whole (\url{https://cubes.inaf.it/home}), and profiles in the main social media, i.e. Facebook, Twitter etc. A series of short video interviews with some of the people in the CUBES consortium have been prepared and made available on the web (\url{https://www.youtube.com/channel/UCZqdt1MnWDUgLcYqjBTFsUA}). As the project becomes more mature, specific activities (conferences, popular science papers, etc.) are foreseen.

\section{CONCLUSIONS}
\label{sec:conclusions}  
We have presented the status of the Cassegrain
U-Band Efficient Spectrograph (CUBES) for the ESO VLT.
Analysis of the design shows that it will deliver outstanding
($>40$\%) throughput across its bandpass (300-405 nm in the present design, exceeding the 305-400 nm TLR), at a mean $R \sim 24000$ (HR mode) and $R \sim 7K$ (LR mode). It is also foreseen an option of a fiber link to UVES to provide the capability of simultaneous high-resolution spectroscopy at $\lambda > 420$ nm.
The CUBES design is fully compliant with the top level requirements and is able to address a treasure trove of scientific cases from Solar System science to Cosmology, with no obvious technical showstopper.
With contributions from institutes in five countries, the CUBES design is well placed to deliver the most
efficient ground-based spectrograph at near-UV wavelengths, with science operations anticipated for 2028, opening unique discovery space for the VLT for years to come.

\acknowledgments

R.S. acknowledges support by the Polish National Science Centre through project 2018/31/B/ST9/01469.
We gratefully acknowledge support from the German Federal Ministry of Education and Research (BMBF) through project 05A20VHA.
B.B. acknowledges the FAPESP grant 2014/18100-4.
For the purpose of open access, the author has applied a Creative Commons Attribution (CC BY) licence to any Author Accepted Manuscript version arising from this submission.

\bibliography{00_CUBES_main} 
\bibliographystyle{spiebib} 

\end{document}